\documentclass[a4,usenatbib]{mnras}
\usepackage{amsmath}
\usepackage{amssymb} 
\usepackage{multirow}
\usepackage{textcomp}
\usepackage{microtype}
\usepackage{subcaption}
\usepackage[dvipsnames]{xcolor}
\usepackage{ulem}
\usepackage{graphicx}
\usepackage{float}
\usepackage{mathptmx}
\usepackage{tabularx}
\usepackage{caption}

\newcommand{\email}[1]{\mbox{\href{mailto:#1}{#1}}}

\def\apj{Astrophysical Journal}                
\def\apjl{Astrophysical Journal}             

\def\apjs{ApJS}              
\def\mnras{MNRAS}            
\def\aap{A\&A}                

\title[Comparing pulsars like clocks]{Applying clock comparison methods to pulsar timing observations}
\author[Chen et al.]{
Siyuan Chen$^{1,2,3}$\thanks{E-mail: \email{siyuan.chen@cnrs-orleans.fr}},
Fran\c{c}ois Vernotte$^{2,4}$
and Enrico Rubiola$^{2,4,5}$
\\
$^1$ Station de Radioastronomie de Nan\c{c}ay, Observatoire de Paris, PSL University, CNRS, Universit\'{e} d'Orl\'{e}ans, 18330 Nan\c{c}ay, France \\
$^2$ FEMTO-ST, Department of Time and Frequency, UBFC and CNRS, 25030 Besan\c{c}on, France \\
$^3$ Laboratoire de Physique et Chimie de l'Environnement et de l'Espace, LPC2E UMR7328, Universit\'{e} d'Orl\'{e}ans, CNRS, 45071 Orl\'{e}ans, France \\
$^4$ Observatoire des Sciences de l'Univers THETA, UBFC and CNRS, 20510 Besan\c{c}on, France \\
$^5$ Istituto Nazionale di Ricerca Metrologica INRiM, Divsion of Quantum Metrology and Nanotechnology, 10135 Torino, Italy
}

\date{Accepted \dots Received \dots; in original form \dots}

\pubyear{2020}

\begin{document}
\label{firstpage}
\pagerange{\pageref{firstpage}--\pageref{lastpage}}
\maketitle

\begin{abstract}

Frequency metrology outperforms any other branch of metrology in accuracy (parts in $10^{-16}$) and small fluctuations ($<10^{-17}$). In turn, among celestial bodies, the rotation speed of millisecond pulsars (MSP) is by far the most stable ($<10^{-18}$). Therefore, the precise measurement of the time of arrival (TOA) of pulsar signals is expected to disclose information about cosmological phenomena, and to enlarge our astrophysical knowledge. Related to this topic, Pulsar Timing Array (PTA) projects have been developed and operated for the last decades. The TOAs from a pulsar can be affected by local emission and environmental effects, in the direction of the propagation through the interstellar medium or universally by gravitational waves from super massive black hole binaries. These effects (signals) can manifest as a low-frequency fluctuation over time, phenomenologically similar to a red noise. While the remaining pulsar intrinsic and instrumental background (noise) are white. This article focuses on the frequency metrology of pulsars. From our standpoint, the pulsar is an accurate clock, to be measured simultaneously with several telescopes in order to reject the uncorrelated white noise. We apply the modern statistical methods of time-and-frequency metrology to simulated pulsar data, and we show the detection limit of the correlated red noise signal between telescopes.

\end{abstract}




\section{Introduction}
\label{sec:intro}

Millisecond pulsars (MSP) are considered extremely stable astronomical clocks because of their high energy and momentum in a small size \citep{2009MNRAS.400..951V}, albeit the observations can be affected by large observational white noise.  Such noise can be due to the low signal to noise ratio of the MSP observations, and depends on the radio-telescope (RT).  On the other hand, red noise could originate from the MSP, the propagation through the interstellar medium or from gravitational waves (GW) on the line of sight, hence it is common to all the RTs \citep{1983ApJ...265L..39H,2015AmJPh..83..635J}.  The characterization of the spectral signature of GWs is a major scientific challenge because it is expected to disclose information about the astrophysical sources.  As pulsars have been precisely timed for decades with up to weekly cadence, the frequency range that can be probed by Pulsar Timing Array (PTA) \citep{1990ApJ...361..300F} projects is $f \in [10^{-9},10^{-6}] \ {\rm Hz}$, which is a window out of reach for the LIGO/VIRGO interferometers and at the edge of the LISA band.  The most interesting and likely to be detected source is a cosmic population of super massive black hole binaries (\cite{2016ApJ...819L...6T} and \cite{2018ASSL..457...95P} for a recent review), whose interaction with the MSP signals introduces a correlated red noise in the Time of arrival (TOA) series, with a phase power spectral density (PSD) proportional to $1/f^{13/3}$ (for a circular population, see eg. \cite{2001astro.ph..8028P,2017MNRAS.470.1738C}).  The choice of the algorithm is therefore of paramount importance to disentangle the pulsar specific small red noise (signal) from the effects of GWs in the presence of large white noise (background) in the shortest observation time.  In this regard, PTAs \citep{2016MNRAS.458.3341D,2020arXiv200506490A,2020PASA...37...20K,2019MNRAS.490.4666P} enable the simultaneous measurement of the same pulsar with several RTs, especially the Large European Array for Pulsars (LEAP) project \citep{2016MNRAS.456.2196B}.

The extraction of small signals from noise exploiting simultaneous measurements is a well-known method for the measurement of oscillators and atomic clocks.  Specific tools have been developed over more than 50 years using the PSD and wavelet variances \citep{Barnes-1971,Rutman-1978,Gray-1974,Allan-1981}.  Among them, the cross-spectrum method \citep{2010arXiv1003.0113R} deserves separate mention in view of our application.  Under certain assumptions, this method rejects the background noise (uncorrelated), and converges to the oscillator noise even if it is significantly lower than the background.  Wavelet variances are commonly used in the time domain, the most known of which is the 2-sample Allan variance (AVAR), after the pioneering work of Allan and Barnes \citep{Allan-1966}.  Such variances enable to distinguish different types of noise defined by their exponent in the PSD, e.g. $f^0$ for white phase, $1/f$ for flicker phase modulation, $1/f^2$ for white frequency modulation, $1/f^3$ for flicker frequency modulation, $1/f^4$ for random walk frequency modulation noise, etc.  The parabolic variance (PVAR) \citep{Benkler-2015,Vernotte-2016} extends this concept, and exhibits (i) the highest rejection of white noise, and (ii) the efficient detection red noise underneath the background with the shortest data record.  Moreover, the wavelet covariance \citep{Fest-1983,Lantz-2019}, i.e., Allan covariance (ACOV) or parabolic covariance (PCOV), enables the rejection of the background using two (or more) uncorrelated instruments. 

This article is intended to port the time-and-frequency metrology methods to pulsar astronomy.  In this respect, the pulsar is the clock under test, observed simultaneously with two or more RTs playing the role of phase meters.  We compare the results of different methods applied to simulated time series representing the TOAs of millisecond pulsars.  In section \ref{sec:intro} we summarize the concept of pulsar timing as well as the spectral and variance methods from the time-and-frequency metrology.  Section \ref{sec:sim} describes the simulation process of the TOA time series, followed by the statistical methods in section \ref{sec:methods}.  Results and the conclusions are presented in sections \ref{sec:results} and \ref{sec:conclusions} respectively.

\subsection{Pulsar timing}

Pulsars are spinning neutron stars that emit radio waves from the region above their magnetic poles. If these regions are misaligned with the rotation axis and happen to point towards Earth as they sweep space, we receive one radio pulse each rotation. The most stable pulsars have rotation period of milliseconds and are stable on a very long timescale. Over the decades that they have been timed, very small variations have been detected, which translate into low frequency red noise in the Fourier domain. A detailed review about pulsars can be found in \cite{2012hpa..book.....L}.

In general, the time of arrival (TOA) of the pulses can be described precisely by the following timing model \citep{2006MNRAS.369..655H}
\begin{equation}
t_{obs} = t_{model} + t_{red} + t_{white} ,
\end{equation}
where $t_{obs}$ is the observed and $t_{model}$ is the predicted TOA considering all known pulsar properties and propagation effects. For this study we also include the effects of dispersion measure variation over time into the perfect model. Ideally, we should be able to precisely predict the TOAs if all model parameters are perfectly known and there are no unknown sources of noise left. However, this is not the case in practice. The difference between the observed and model observation forms the residual series $x=t_{red} + t_{white}$. Therefore, we are interested in both white and red noise components.

The pulse residual series can be transformed into Fourier frequency space. The noise components are then described by a PSD. The PTA noise description can be found in \cite{2016MNRAS.458.2161L,2016MNRAS.457.4421C,2016ApJ...821...13A}. Following the notation of the latter, the white noise is described by two parameters $E_k$ and $Q_k$
\begin{equation}
S_{w} = (E_k W)^2 + Q_k^2 ,
\label{eqn:PTA_white}
\end{equation}
where $W$ is the initial estimate of the white coming from the radiometer noise and template matching error, $E_k$ and $Q_k$ are the so-called EFAC and EQUAD parameters respectively. Note that $S_{w}$ is constant over the whole frequency range.

The red noise is described by a power law with two parameters
\begin{equation}
S_{r}(f) = A_{PTA}^2 f^{-\gamma_{PTA}} ,
\end{equation}
where $A_{PTA}$ is the amplitude and $\gamma_{PTA}$ is the spectral index of the power law. The total noise PSD $S(f)$ becomes \citep{808879}
\begin{equation}
S(f) = S_{r}(f) + S_{w} .
\label{eqn:PTA_SD}
\end{equation}

\subsection{Power and Cross spectrum}

\begin{figure}
\centering
\includegraphics[width=0.9\linewidth]{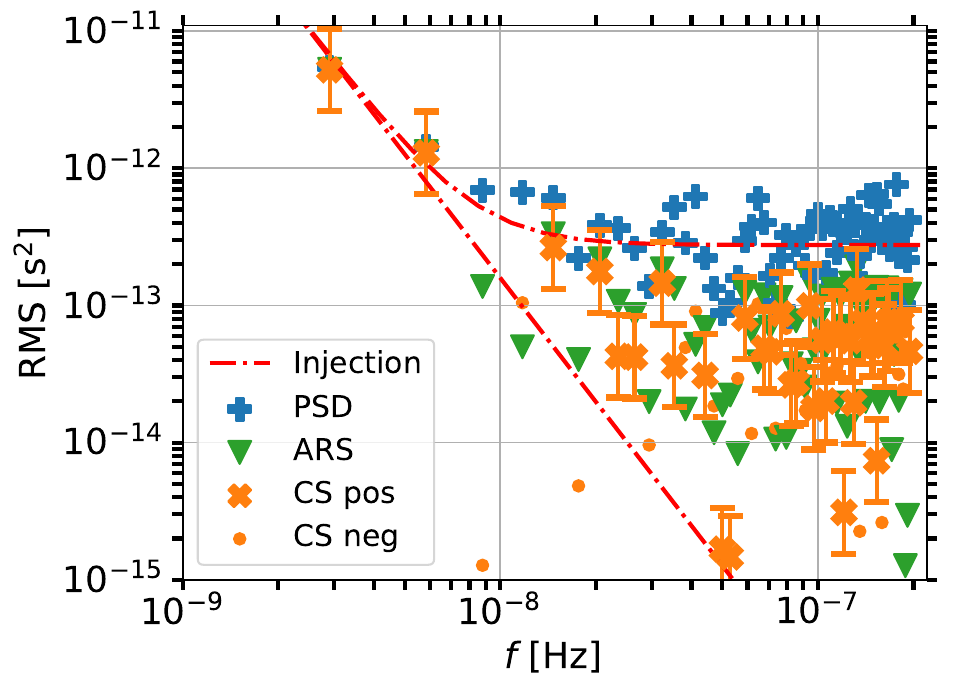}
\caption{Average PSD (blue pluses), averaged residual spectrum (ARS) (green triangles) and CS (orange crosses/points) computed from one set of 5 different residual series in the boundary case (red dashed lines)}
\label{fig:SD}
\end{figure}

In time and frequency metrology, the cross spectrum is the most often used method to measure the phase noise of oscillators. The method relies on the simultaneous measurement with two separate instruments, assuming that the background noise is statistically independent. The same hypothesis holds for pulsar observations from different radio telescopes. The telescopes are obviously independent, and sufficiently far from each other for the tropospheric/ionospheric noise to be statistically independent. We summarize the main results from our tutorial on the cross spectrum method \citep{2010arXiv1003.0113R}.

Given a residual series $x$ from an instrument, we can write its Fourier transform as $X = \Re(X) + i \: \Im(X)$. 
The one-sided PSD can be calculated as
\begin{equation}
S_x(f) = \frac{2}{T_a} X X^* = \frac{2}{T_a} \Big( |\Re(X)|^2 + |\Im(X)|^2 \Big) \qquad\text{for $f>0$} ,
\label{eqn:PSD}
\end{equation}
where $T_a$ is the acquisition time, and the factor 2 is due to energy conservation after surpressing the negative frequencies.

Measuring a clock (pulsar) signal $c$ simultaneously with two instruments (telescopes), we get two residual series $x$ and $y$, each containing the white background noise of the measurement added to $c$. The cross spectrum (CS)
\begin{equation}
S_{yx} = \frac{2}{T_a} YX^*
\end{equation}
converges to $S_c = \frac{2}{T_a} CC^*$. Notice that, after compensating for the differential path, $c$ goes only into $\Re(S_{yx})$, while the $\Im(S_{yx})$ contains only the background noise of the observation. After \cite{2010arXiv1003.0113R}, the fastest, unbiased estimator is
\begin{equation}
\widehat{S_c} = \frac{2}{T_a} \Re(S_{yx}) = \frac{2}{T_a} \Big( \Re(X) \: \Re(Y) + \Im(X) \: \Im(Y) \Big)
\label{eqn:CS}
\end{equation}
and can give negative values. We average over all the CSs coming from the different RT pairs to produce one simultaneous observation series for the analysis.

Figure \ref{fig:SD} shows an example of a comparison between the different spectra. The PSD can be computed on the averaged residual series (ARS) or on each residual series individually, and then averaged (PSD). As the ARS is very similar, but slightly worse than the CS, we opt to focus the analysis on the CS method, whose results should be comparable to those from the ARS. The CS is better at rejecting uncorrelated white noise and produces a lower level than the average PSD, which serves as a good reference.

We also investigated the effects of the choice of the window function on the Fourier transform by comparing the sinc function used above with a Blackman window. Although there is spectral leakage of the red noise, the amount is only noticable in the lowest frequencies and is within the uncertainty of the PSD/CS value. We found minimal effects on our analysis. In fact, as the spectral value decreases, the recovered red noise parameters provide a worse match to the injected values. We also found similar results with a prewhitening / postdarkening procedure. Therefore, we show only the results obtained with the above equations.

\subsection{Variances}

\begin{figure}
\centering
\includegraphics[width=0.9\linewidth]{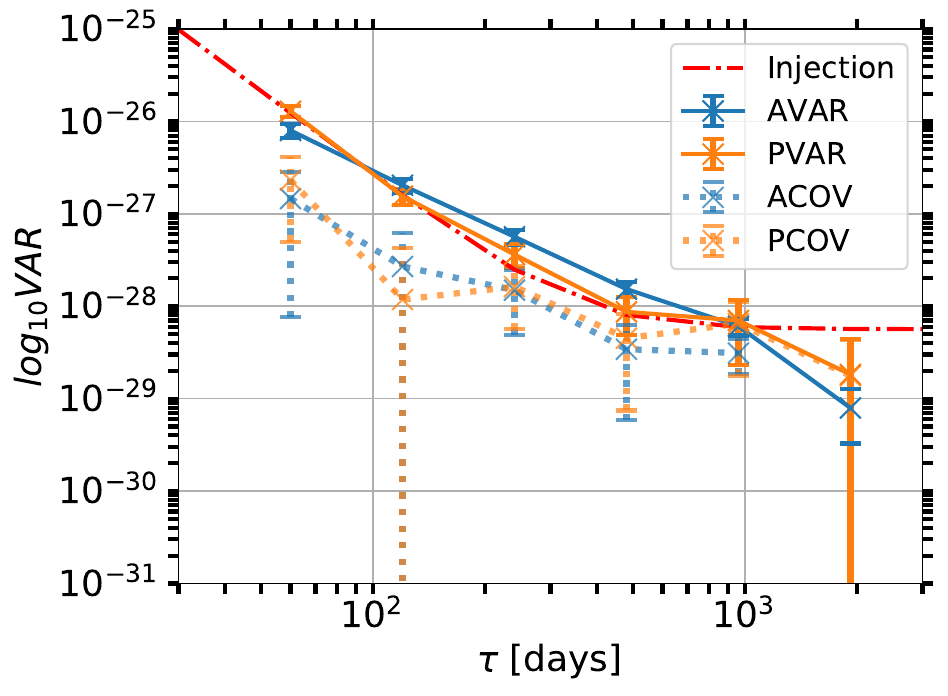}
\caption{Averaged data Allan variance (blue) and parabolic variance (orange), the corresponding covariances (light dotted lines) and the associated uncertainties computed from one set of 5 residual series in the boundary case (red dashed line)}
\label{fig:VAR}
\end{figure}

Different variances have been proposed to describe the amount of variation in a given residual series $x$ with a total time span $T$ for time steps $\tau$. The two prominent ones are the Allan \citep{Allan-1966} and more recently the parabolic variance \citep{Vernotte-2016}, which we summarize here.

The AVAR is a simple computation comparing three observations separated by $m$ time steps
\begin{equation}
{\rm AVAR} (\tau) = \frac{1}{2 M \tau^2} \sum_{i=0}^{M-1} \Big[ (-x_{i} + 2x_{i+m} -x_{i+2m}) \Big]^2 ,
\label{eqn:AVAR}
\end{equation}
where $M = N-2m+1$ is related to the total number $N$ of data points in the residual series $x$. Compared to similar variances, AVAR is the most efficient at estimating for the largest $\tau = T/2$. By contrast, the Modified Allan variance is more suitable to the analysis of short-term fluctuations \citep{Allan-1981}. The parabolic variance combines the original long-term computability as well as a good response to short timescales
\begin{equation}
{\rm PVAR} (\tau) = \frac{72}{M m^4 \tau^2} \sum_{i=0}^{M-1} \Big[ \sum_{k=0}^{m-1} (\frac{m-1-2k}{2}) (x_{i+k} - x_{i+m+k}) \Big]^2 ,
\label{eqn:PVAR}
\end{equation}
where $M = N-2m+2$.

Similarly to the CS method, we can also use two indepedent residual series $x$ and $y$ to compute the ACOV
\begin{equation}
\begin{split}
{\rm ACOV} (\tau) = \frac{1}{2 M \tau^2} \sum_{i=0}^{M-1} \Big[ & (-x_{i} + 2x_{i+m} -x_{i+2m})
\\
& (-y_{i} + 2y_{i+m} -y_{i+2m}) \Big]
\end{split}
\label{eqn:ACOV}
\end{equation}
and the PCOV
\begin{equation}
\begin{split}
{\rm PCOV} (\tau) = \frac{72}{M m^4 \tau^2} \sum_{i=0}^{M-1} \Big[ \sum_{k=0}^{m-1} & (\frac{m-1-2k}{2}) (x_{i+k} - x_{i+m+k})
\\
& (\frac{m-1-2k}{2}) (y_{i+k} - y_{i+m+k}) \Big] .
\end{split}
\label{eqn:PCOV}
\end{equation}

A comparison of the two different (co)variances can be found in figure \ref{fig:VAR}. In general, the variance plot has two prominent features: (i) at lower time steps, the variance is dominated by white noise, which leads to a power law decrease with fixed spectral index, (ii) at higher time steps, red noise start to appear, which shows up as a flattening and then rising of the variance, depending on the amplitude and spectral index of the red noise. The variance curve can be parametrized as
\begin{equation}
VAR = A_{VAR}^2 \tau^{\gamma_{VAR}} + B_{VAR}^2 \tau^{\beta_{VAR}} .
\label{eqn:PTA_VAR}
\end{equation}
The relation between the PTA parameters EFAC, EQUAD, $\gamma_{PTA}$ and $A_{PTA}$ and the variance parameters is not trivial. As the white noise follows a fixed theoretical spectral index for a given variance $\beta_{VAR}=-2/-3$ for AVAR and PVAR respectively, we can compare the overall amplitude of the variance white noise with $S_{w}$ from the PTA white noise in \eqref{eqn:PTA_white}
\begin{equation}
S_{w} = 
\begin{cases}
2 B_{VAR}^2/3 & \text{for AVAR}
\\
B_{VAR}^2/(6 \tau_0) & \text{for PVAR} ,
\end{cases}
\label{eqn:VAR_white}
\end{equation}
where $\tau_0$ is the sampling time.

The two red noise spectral indices can be related via 
\begin{equation}
\gamma_{PTA} = \gamma_{VAR}+3 .
\label{eqn:VAR_gamma}
\end{equation}
The red noise amplitudes follow analytic relations for given integer spectral indices \citep[see table 1 in][]{Vernotte-2016}. Under the condition that $\gamma_{PTA} \in [1.5,4.5]$ it is also possible to find an analytic relation between $A_{VAR}$ and $A_{PTA}$ for non-integer $\gamma_{PTA}$, see also \cite{2020arXiv200513631V},
\begin{equation}
A_{VAR}^2 = \frac{A_{PTA}^2}{12\pi^2} \: 4\pi^2 \: yr^{3+\gamma_{PTA}} \mathcal{A}(\gamma_{PTA}) ,
\label{eqn:VAR_red}
\end{equation}
where $\mathcal{A}(\gamma_{PTA})$ is a function that relates to the coefficients of the variance. Using the same approach as \cite{286357} we find the folowing relationships for AVAR and PVAR
\begin{equation}
\mathcal{A}(\gamma_{PTA}) = 
\begin{cases}
(4^{\gamma_{PTA}-2} - 2^{\gamma_{PTA}-1}) \: \pi^{\gamma_{PTA}-3} \\ \Gamma(1-\gamma_{PTA}) \sin(\pi-\pi\gamma_{PTA}/2) & \text{for AVAR} \\
\\
-4^{\gamma_{PTA}} \: 9\pi^{\gamma_{PTA}-3} \\ (2^{2-\gamma_{PTA}} (-1-\gamma_{PTA}) +2 +3\gamma_{PTA} -\gamma_{PTA}^2) \\ \Gamma(-3-\gamma_{PTA}) \sin(\pi-\pi\gamma_{PTA}/2) & \text{for PVAR} .
\end{cases}
\label{eqn:VAR_norm}
\end{equation}

\section{Simulated datasets}
\label{sec:sim}

Clock comparison methods require several time series of the same pulsar observed simultaneously with different telescopes over a long time span. The LEAP project uses the 5 major radio telescopes in Europe: Effelsberg in Germany, Lovell in UK, Westerbork in the Netherlands, Nan\c{c}ay in France and Sardinia in Italy as an interferometer with monthly simultaneous observations since 2012 \citep{2016MNRAS.456.2196B,2017A&C....19...66S}.

Thus we simulate 100 realizations of 5 TOA series over a 4000-day long observing campaign with a cadence of 30 days, this is of the order of the nominal LEAP observations up to today. They have been created using the \texttt{libstempo} package \citep{libstempo}. For simplicity, we assume that all 5 RTs have comparable instrumental white noise, and give to all residuals the initial uncertainty $W = 500$ ns. As the two white noise parameters are degenerate, we will only focus on the dominant term, ie. EFAC, injecting different realizations of white noise with (${\rm EFAC}=1.05, {\rm EQUAD} = 0$).

We also simulate the same residual series as it would be observed by LEAP under ideal conditions. In the LEAP statistical limit (LSL) of the coherently added mode, the total S/N of the LEAP observation is the sum of the individual RTs. Since we assume that the 5 RTs are identical, the LEAP observations have a 5 times higher S/N with the TOAs having a white noise level that is 5 times lower as $S/N \propto 1/W$. This also translates into an EFAC injection of $1.05/5=0.21$, while EQUAD remains at 0.

In total, we consider three cases:
\begin{enumerate}
\item white noise case: no red noise signal injected
\item boundary case: ($\log_{10} A_{PTA} = -14, \gamma_{PTA} = 3$)
\item red noise case: ($\log_{10} A_{PTA} = -13, \gamma_{PTA} = 3$)
\end{enumerate}

\begin{figure}
\centering
\includegraphics[width=0.9\linewidth]{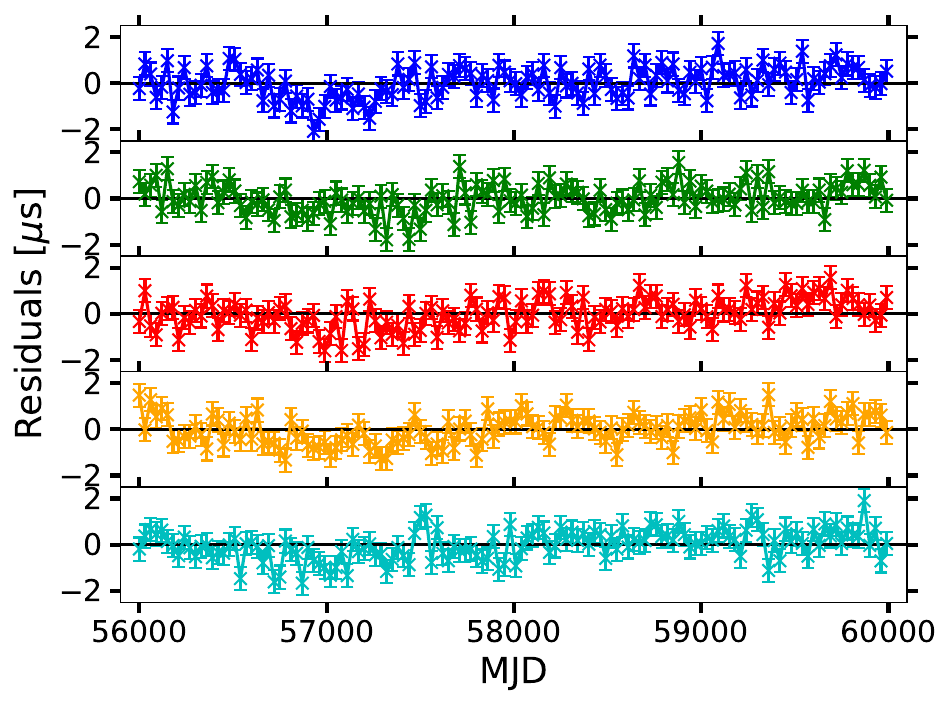}
\caption{Example set of 5 different simulated residual series of a 4000-day long observing campaign with a monthly cadence in the boundary case}
\label{fig:residuals}
\end{figure}

For each case we simulate 100 sets of 5 residual series with different red noise realization between each set. For a given set, the red noise realization is the same amongst the 5 residual series, but the white noise realizations differ. An example of a set of 5 simulated residual series can be seen in figure \ref{fig:residuals}. As we are focusing only on the noise properties, we fix the timing model at its injected values. This means that the residual series are fully determined by the injected white and red noise. In practice, there will be covariances between the timing model and the noise parameters. Here we only show the theoretically optimal results as a proof-of-concept.

\section{Methods}
\label{sec:methods}

The standard PTA analysis is done via the \texttt{enterprise} package \citep{enterprise}, details of the analysis can be found in \cite{2016ApJ...821...13A,2018ApJ...859...47A}. We use all 5 individual RT residual series for the PTA analysis comparison. For the LSL analysis we only have one residual series, but with a lower white noise level.

We compute the PSD and CS through equations \eqref{eqn:PSD} and \eqref{eqn:CS}, see figure \ref{fig:SD}, and compare them directly to equation \eqref{eqn:PTA_SD} to get constraints on the noise parameters. The PSD can be averaged over the 5 residual series, whilst the CS can be computed from the average of the 10 different combinations of 2 out of the 5 residual series.

The Allan and parabolic Variances are calculated through equations \eqref{eqn:AVAR} and \eqref{eqn:PVAR}, see figure \ref{fig:VAR}, and matched to the parametrization in equation \eqref{eqn:PTA_VAR} with the use of equations \eqref{eqn:VAR_white}, \eqref{eqn:VAR_gamma} and \eqref{eqn:VAR_red}. Unlike the PSD, we use the averaged residual series to compute the variances.

We use Bayesian statistics and a nested sampling algorithm, implemented in the \texttt{cpnest} package \citep{cpnest}, to look for the posterior constraints on the white noise properties as well as the spectral index and amplitude of the red noise parameters.

A likelihood function can be derived from a weighted least squares fitting procedure
\begin{equation}
\mathcal{L} = -\frac{1}{2\sigma_i^2} \Big( \mathcal{D} - \mathcal{F}(\phi) \Big)^2 ,
\label{eqn:likelihood}
\end{equation}
where $\sigma_i^2$ are the weights, $\mathcal{D}$ are the computed data points and $\mathcal{F}(\phi)$ are the corresponding parametrizations with the model parameters $\phi = ({\rm EFAC}, \log_{10} {\rm EQUAD}, \gamma_{PTA}, \log_{10} A_{PTA})$.

For the spectral analysis we work in the logarithmic space, where $\mathcal{D}$ corresponds to the computed $\log_{10}$ PSD/CS values. From a large Monte-Carlo simulation we infer the uncertainties to be $\sigma_i = 0.2136$ across all frequencies in the Gaussian approximation using only the positive estimates, see appendix \ref{sec:weights_sd}.

For the (co)variance analysis, where $\mathcal{D}$ are the computed (co)variances, we compute the weights $\sigma_i^2$ from the effective degrees of freedom for each time step. A more detailed description can be found in appendix \ref{sec:weights_var}.

The choice of priors follows closely the standard PTA analysis (eg. \cite{2015MNRAS.453.2576L}) wherever possible with uniform distributions in: ${\rm EFAC} \in [0.1,1.5]$, $\log_{10} {\rm EQUAD} \in [-10,-5]$, $\gamma_{PTA} \in [0,7]$ and $\log_{10} A_{PTA} \in [-20,-10]$. The only exception are the boundaries of $\gamma_{PTA} \in [1.5,4.5]$ for the variance analysis. This is due to the limited range of validity of equation \eqref{eqn:VAR_norm} \citep{2020arXiv200513631V}.


\section{Results}
\label{sec:results}

To ensure that our results are statistically robust, each of the three cases is based on 100 different realization sets and doing the analysis using all the clock comparison methods and the standard PTA/LSL comparisons. All following posterior distributions and recoveries are based on the addition of all 100 individual posterior distributions, unless stated otherwise. They should thus be representative of the constraints from the analyses of a single simulated set of 5 residual series.

\subsection{White noise}

\begin{figure}
\centering
\includegraphics[width=0.9\linewidth]{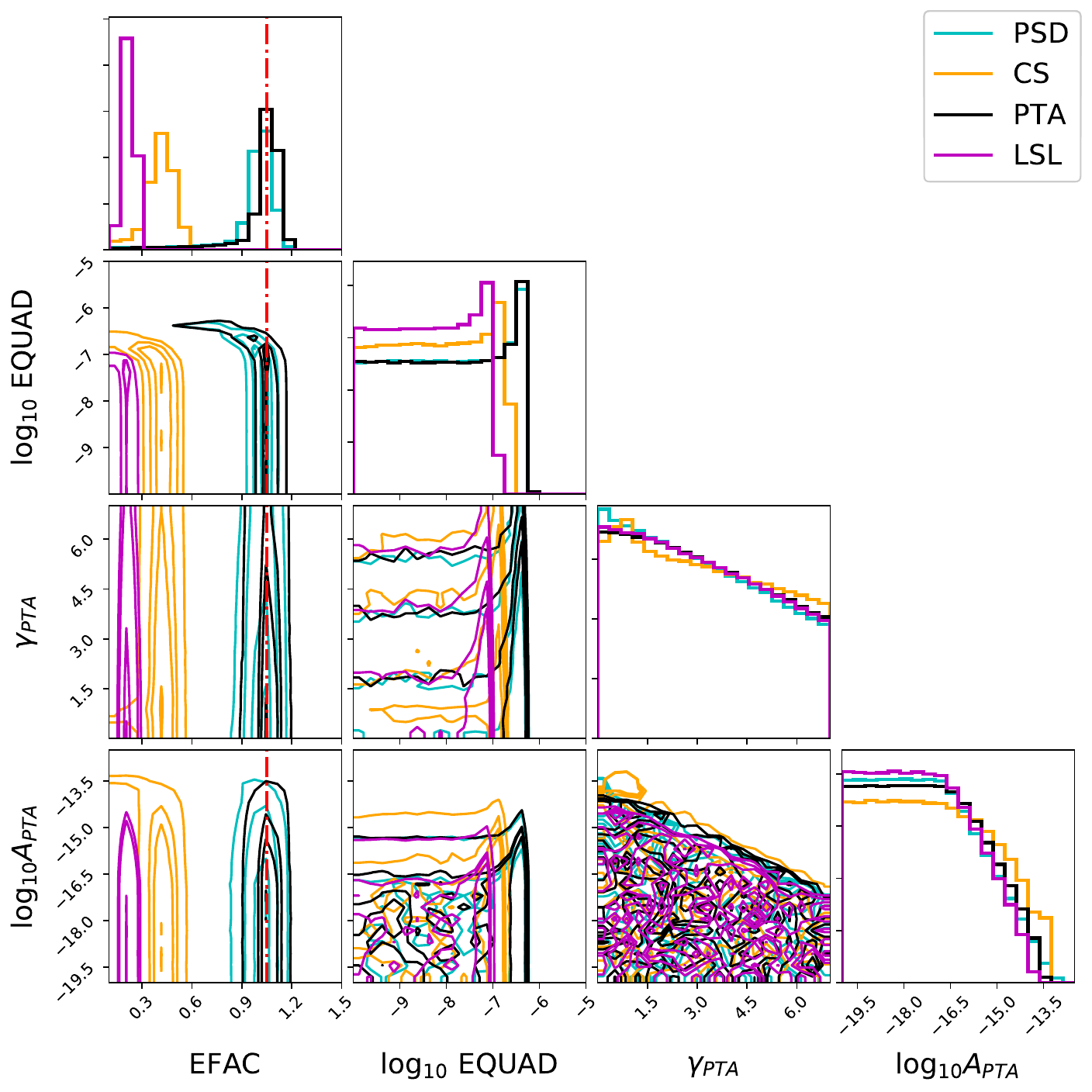}\\
\includegraphics[width=0.9\linewidth]{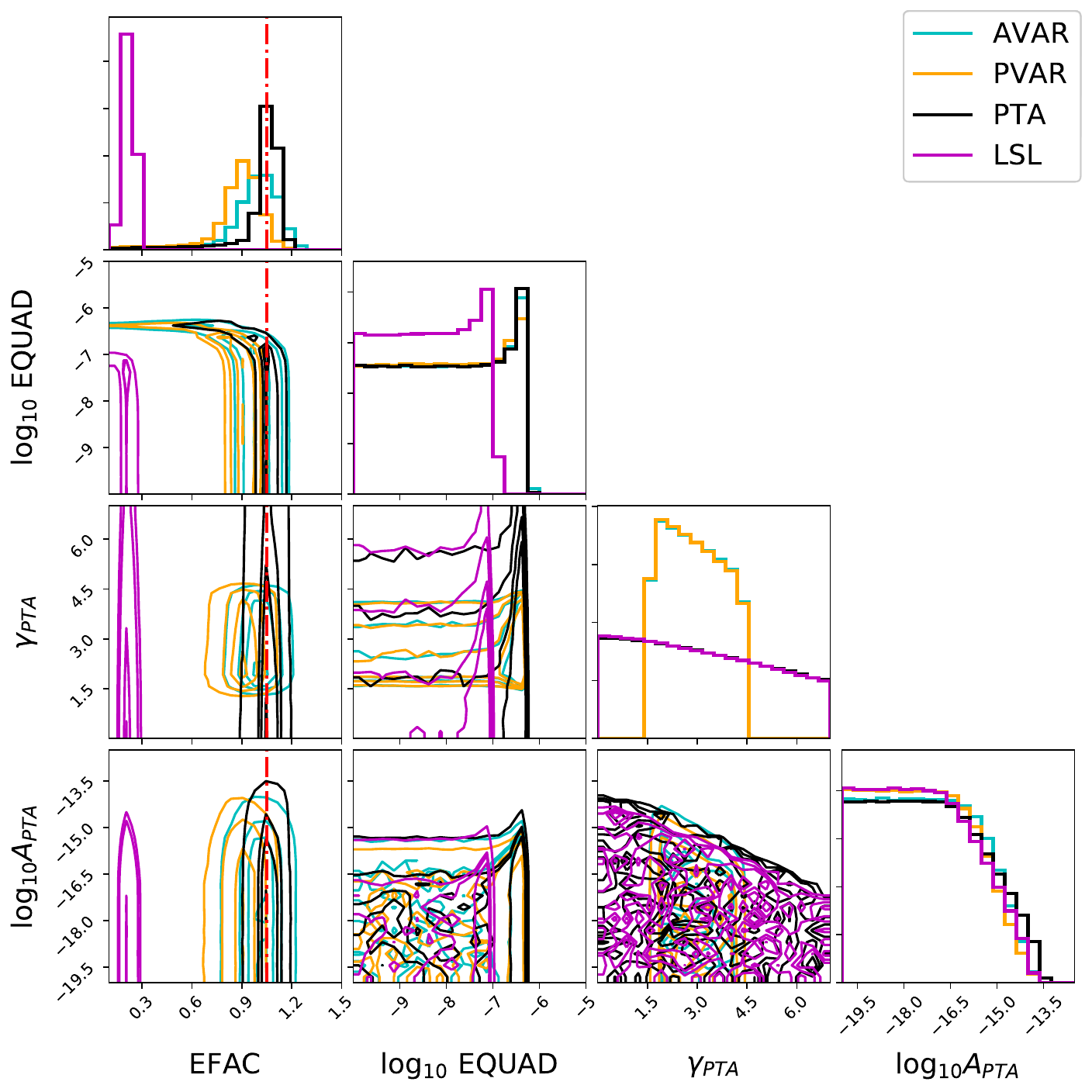}\\
\includegraphics[width=0.9\linewidth]{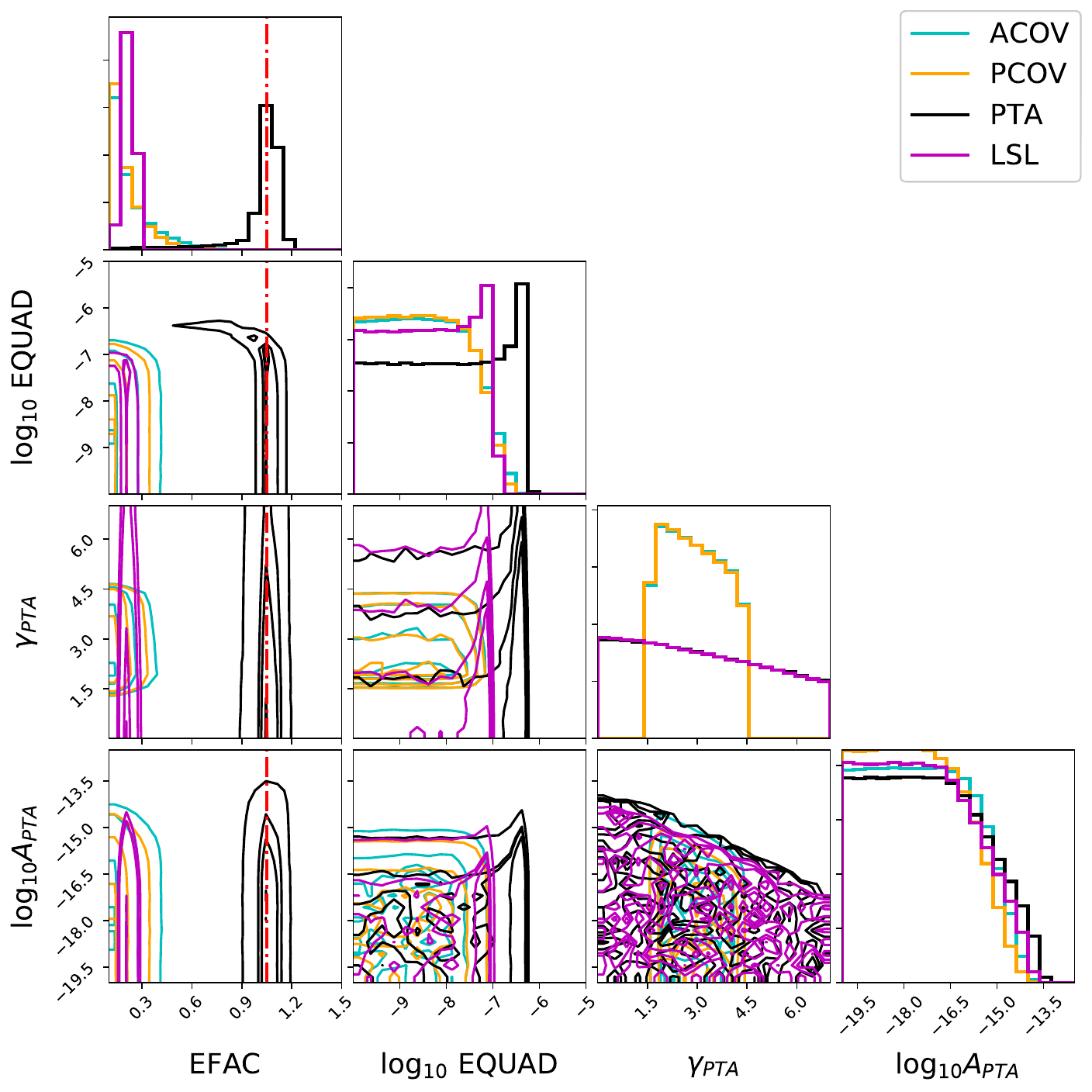}
\caption{Comparison of the posterior distribution corner plots for the white noise case: spectral (top), variance (middle) and covariance analyses (bottom)}
\label{fig:white_corner}
\end{figure}

We first explore the performance of the clock methods in the case of pure white noise in the residual series, and compare it to the traditional PTA/LSL analysis. Figure \ref{fig:white_corner} shows the PTA noise parameter marginalized 1D and 2D posterior distributions in a corner plot. As there is no red noise present, $A_{PTA}$ and $\gamma_{PTA}$ return very uninformative posteriors for all methods. $\gamma_{PTA}$ simply returns a nearly flat uniform prior. The quantity $A_{PTA}$ returns a posterior distribution that is flat for low amplitudes, but has a cutoff at an higher amplitude. This can be interpreted as a upper limit case, as red noise with a higher amplitude should have affected the residual series. Similarly, as there is no EQUAD injected, all analyses also show upper limits.

The only parameter of interest in the case of white noise is EFAC. As expected, both the PSD and PTA methods return distributions around the injected value, with the PTA distribution being more constraining than the PSD. The CS method does not seem to lower the white noise to the same level as the LSL, see top panel of figure \ref{fig:white_corner}. However, this is misleading, as the white noise is constrained from the positive CS values only. In fact, the CS method lowers the white noise to almost zero and could perform similarly or better to the LSL residuals by also taking the negative CS values into account. This would, however, come at the cost of a decreased performance in the recovery of the red noise parameters. In the middle panel of figure \ref{fig:white_corner}, both AVAR and PVAR show similar, but worse, contraints than the PTA analysis, where AVAR is closer to the injected value and PVAR has a small bias towards lower values. The bottom panel shows that the covariances reject most of the white noise, pushing the EFAC towards to lower end of the allowed values. This effect is intended as the covariances `try' to find a correlated signal, but as there is only uncorrelated noise a upper limit on EFAC is placed.

\begin{figure*}

\includegraphics[width=0.3\linewidth]{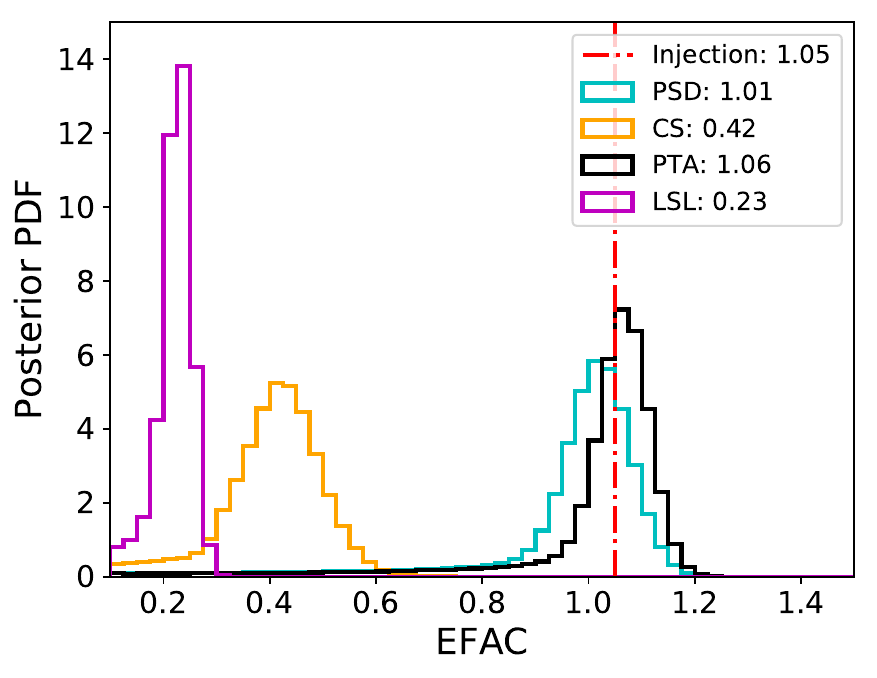} \hspace{0.5cm}
\includegraphics[width=0.3\linewidth]{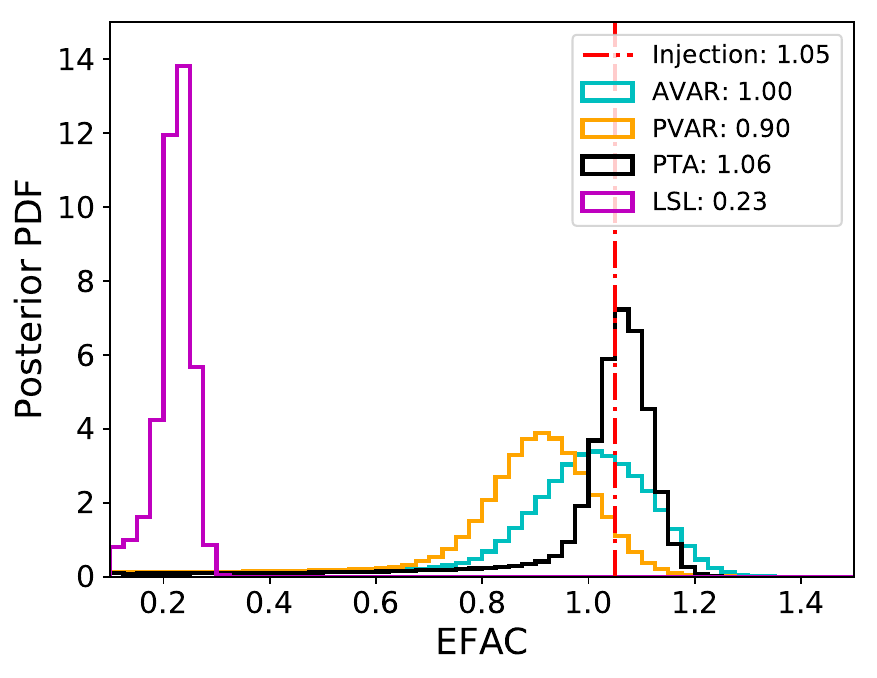} \hspace{0.5cm}
\includegraphics[width=0.3\linewidth]{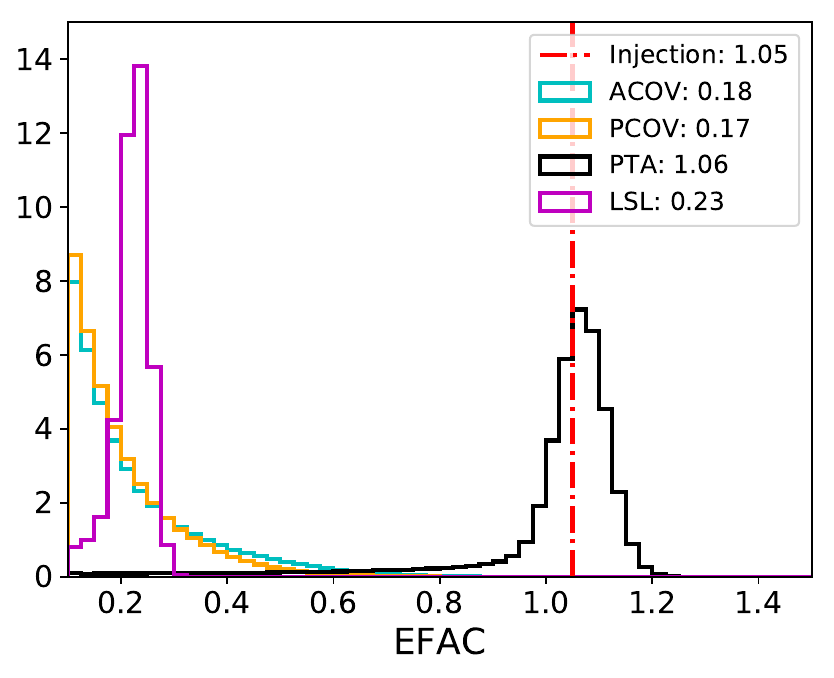}
\\
\includegraphics[width=0.3\linewidth]{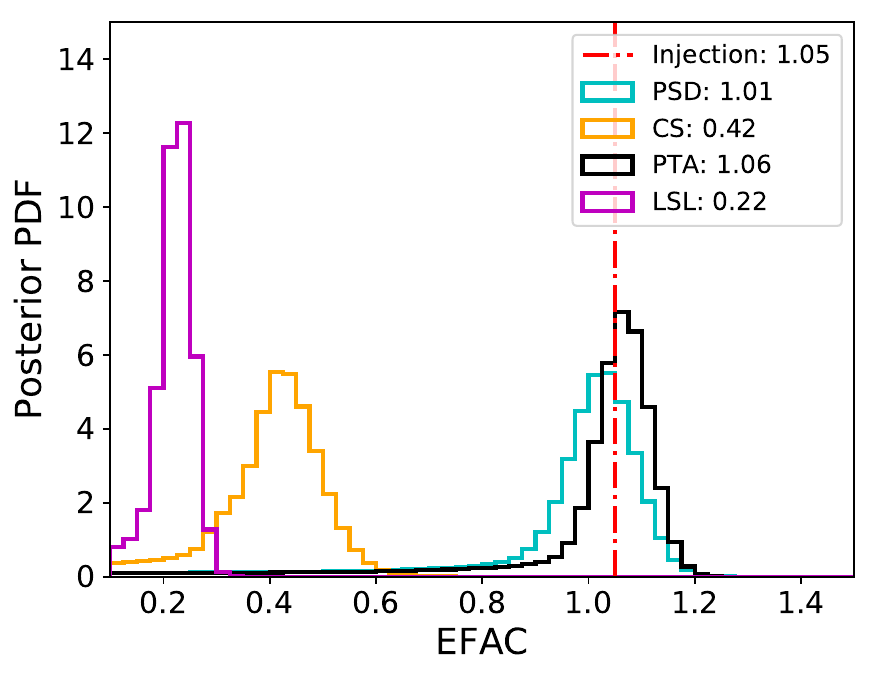} \hspace{0.5cm}
\includegraphics[width=0.3\linewidth]{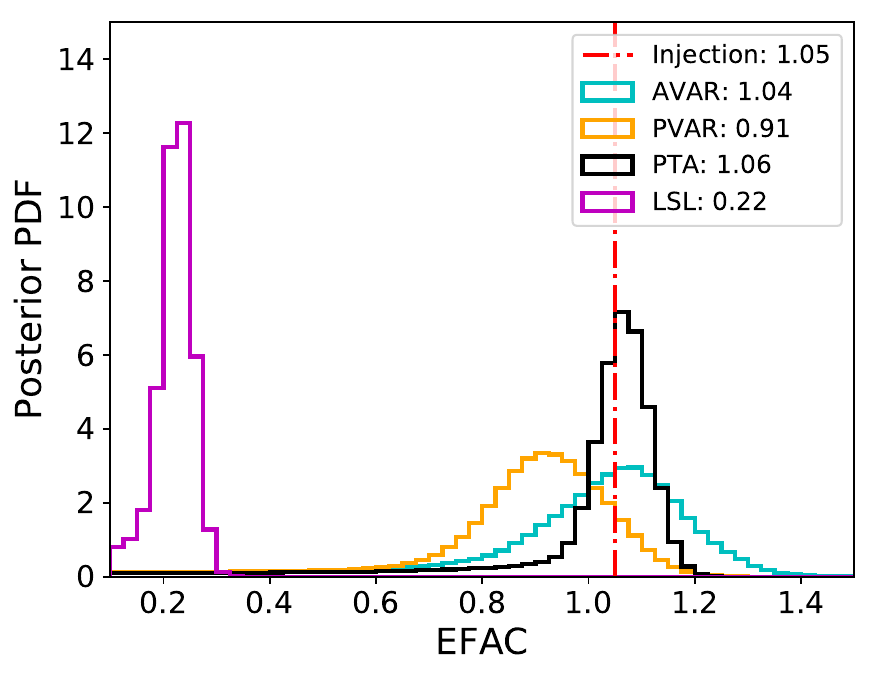} \hspace{0.5cm}
\includegraphics[width=0.3\linewidth]{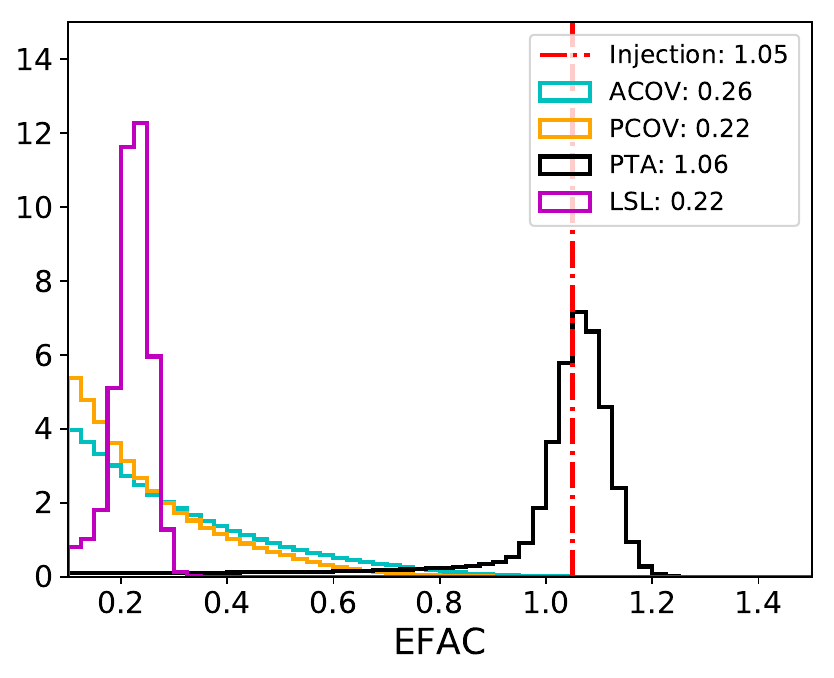}
\\
\includegraphics[width=0.3\linewidth]{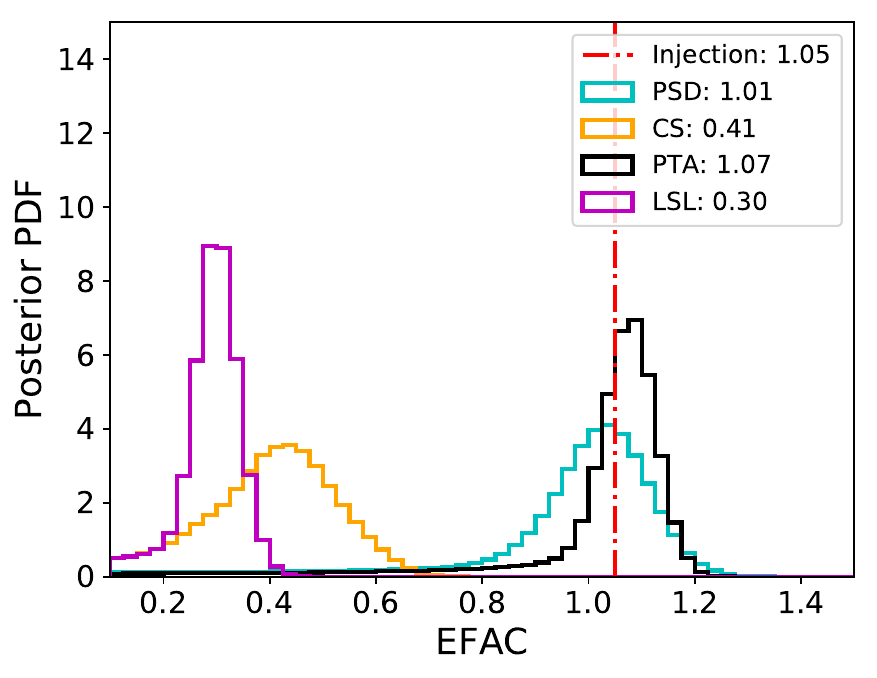} \hspace{0.5cm}
\includegraphics[width=0.3\linewidth]{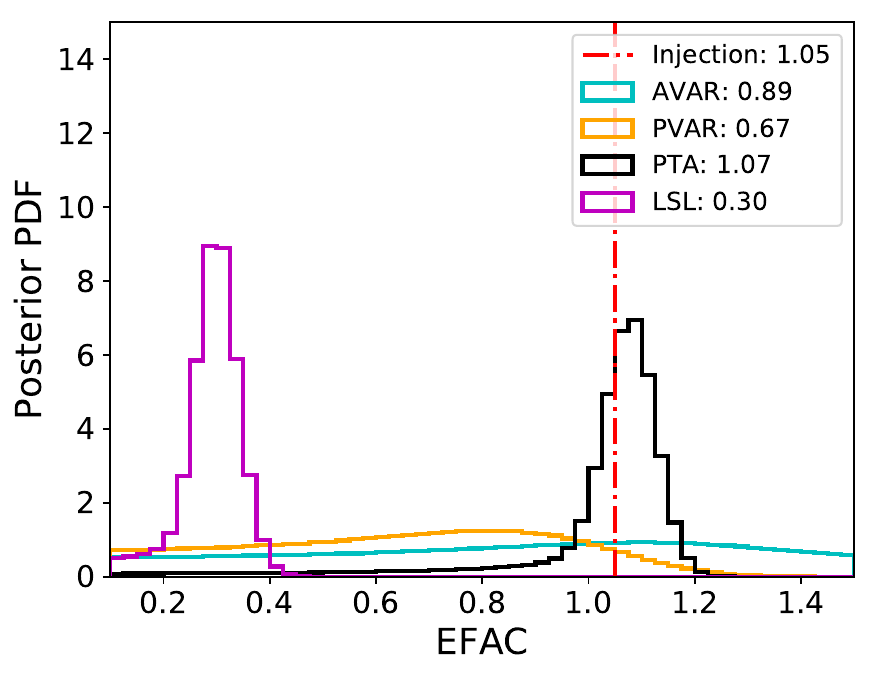} \hspace{0.5cm}
\includegraphics[width=0.3\linewidth]{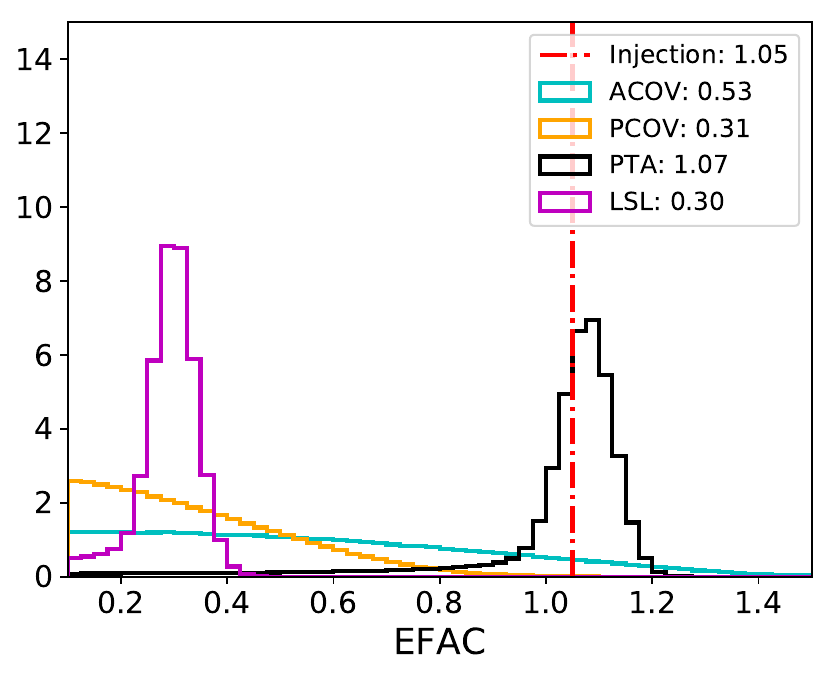}

\caption{Evolution of the recovery of the EFAC parameter from the white noise case (top row), boundary case (middle row) to the red noise case (bottom row): spectral (left column), variance (middle column) and covariance methods (right column). The numbers in the legends represent the median values of the posterior distributions.}
\label{fig:white_evolution}
\end{figure*}

However, we track the evolution of the performance of the different methods to recover white noise as the amount of red noise gradually increases as we move through the 3 simulated cases. The main result in figure \ref{fig:white_evolution} is that as the red noise increases, all methods seem to progressively perform worse in constraining the white noise. For the spectral and variance analysis, the posterior distributions progressively widen. Particularly in the strong red noise case, the AVAR and PVAR analysis are completely unable to recover the injected white noise (bottom middle panel). For the covariance analysis, the EFAC upper limit gradually increases, most notable in the strong red noise case.

\begin{figure*}

\includegraphics[width=0.3\linewidth]{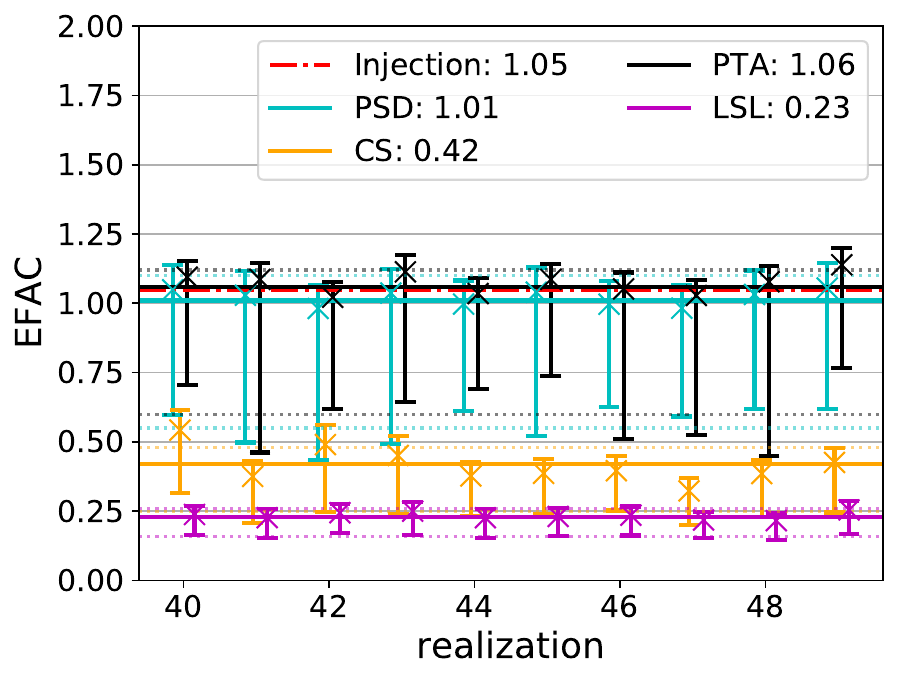} \hspace{0.5cm}
\includegraphics[width=0.3\linewidth]{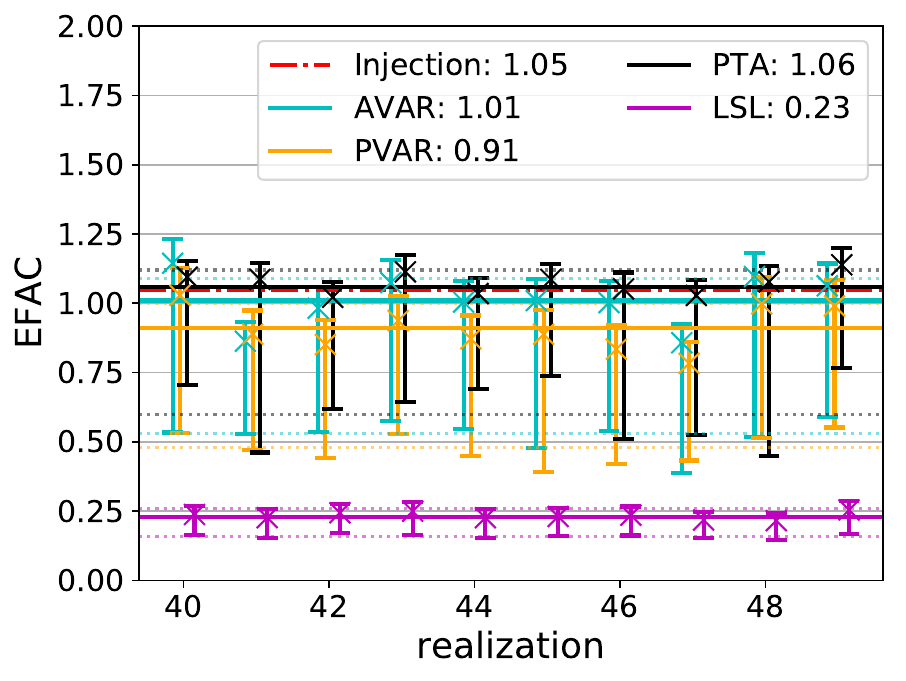} \hspace{0.5cm}
\includegraphics[width=0.3\linewidth]{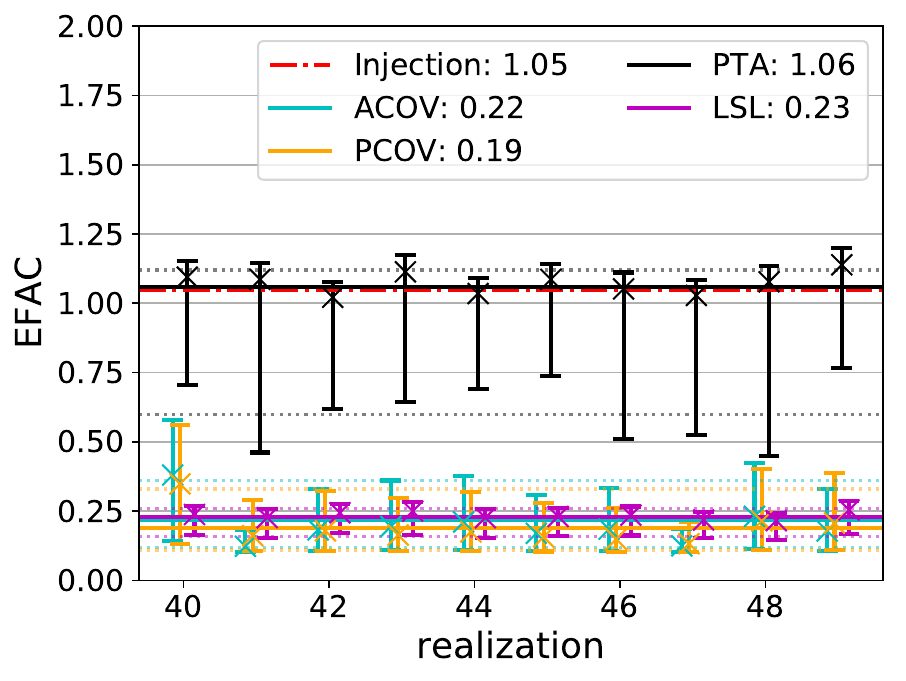}
\\
\includegraphics[width=0.3\linewidth]{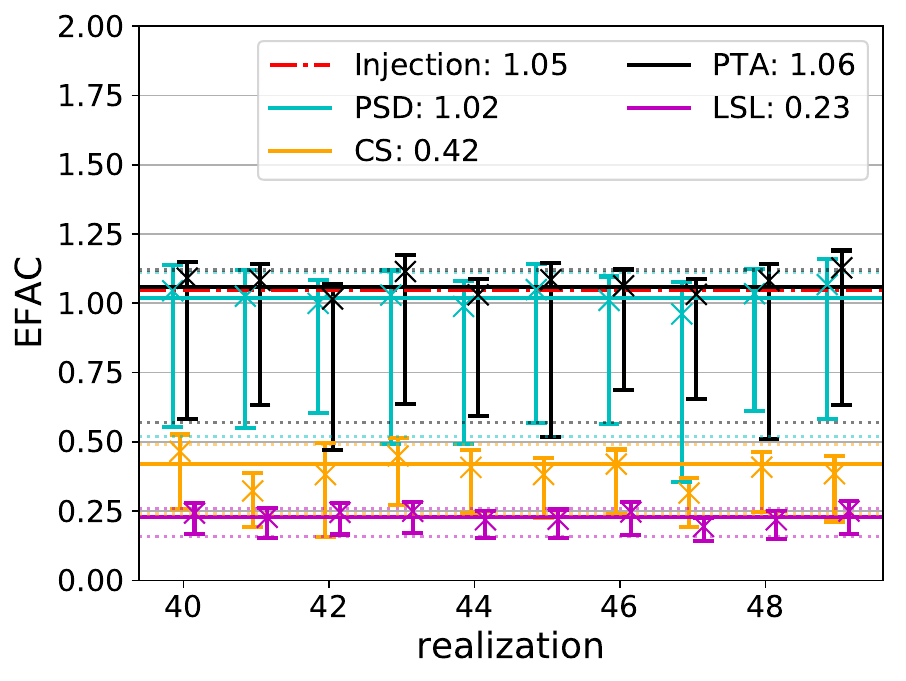} \hspace{0.5cm}
\includegraphics[width=0.3\linewidth]{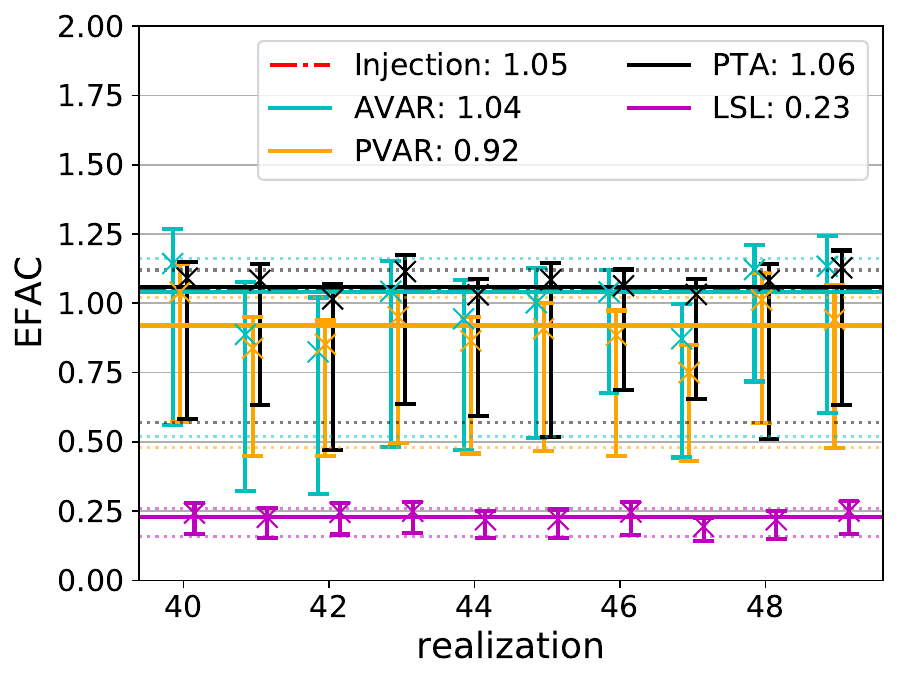} \hspace{0.5cm}
\includegraphics[width=0.3\linewidth]{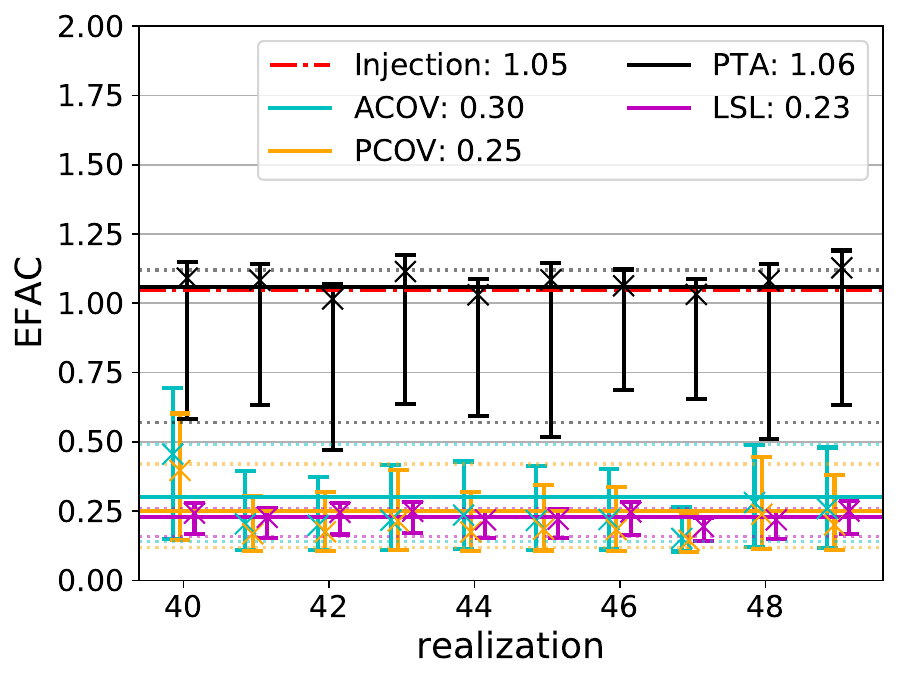}
\\
\includegraphics[width=0.3\linewidth]{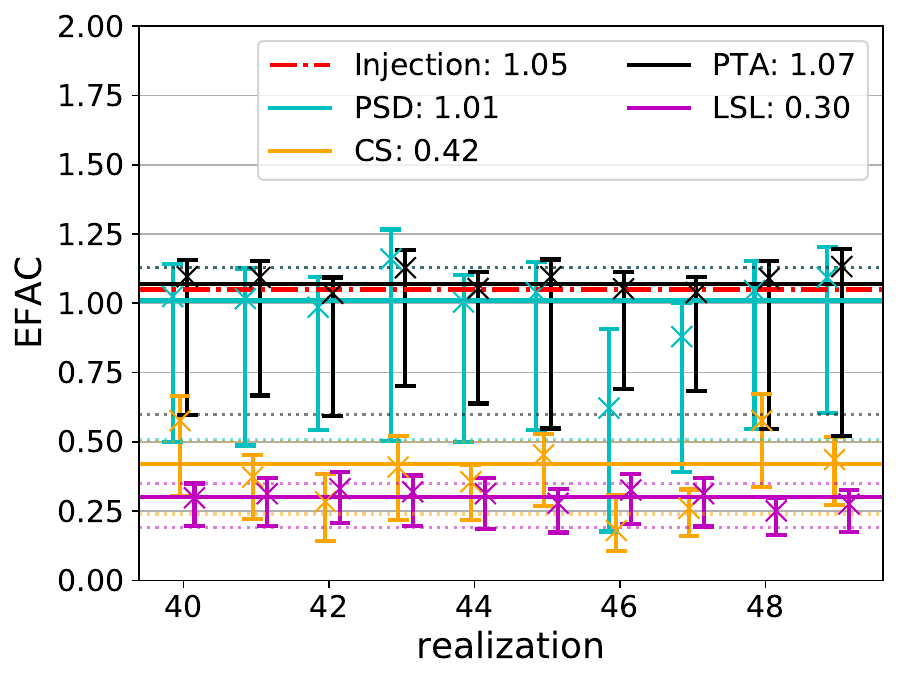} \hspace{0.5cm}
\includegraphics[width=0.3\linewidth]{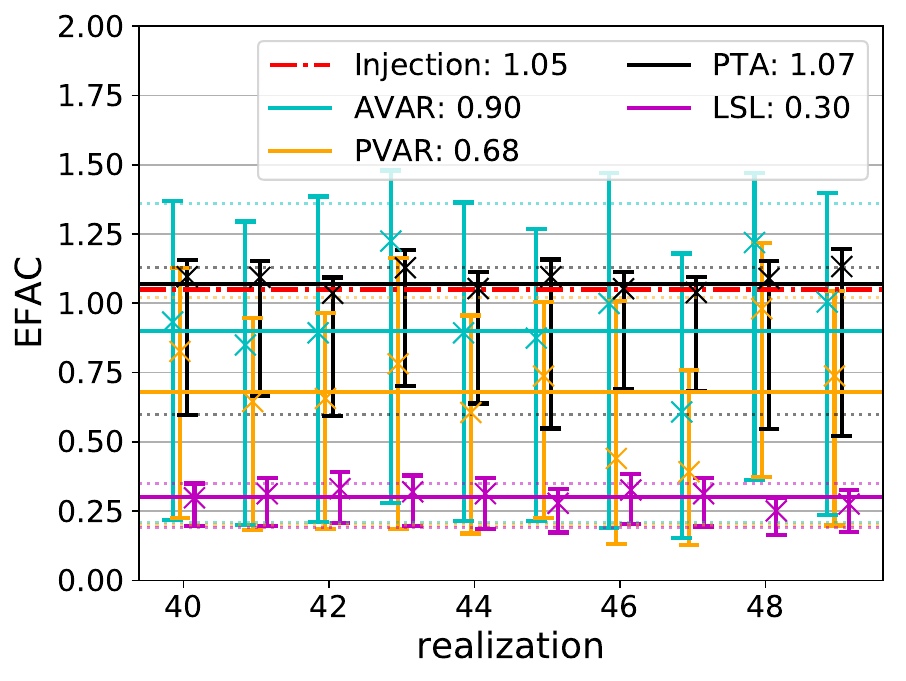} \hspace{0.5cm}
\includegraphics[width=0.3\linewidth]{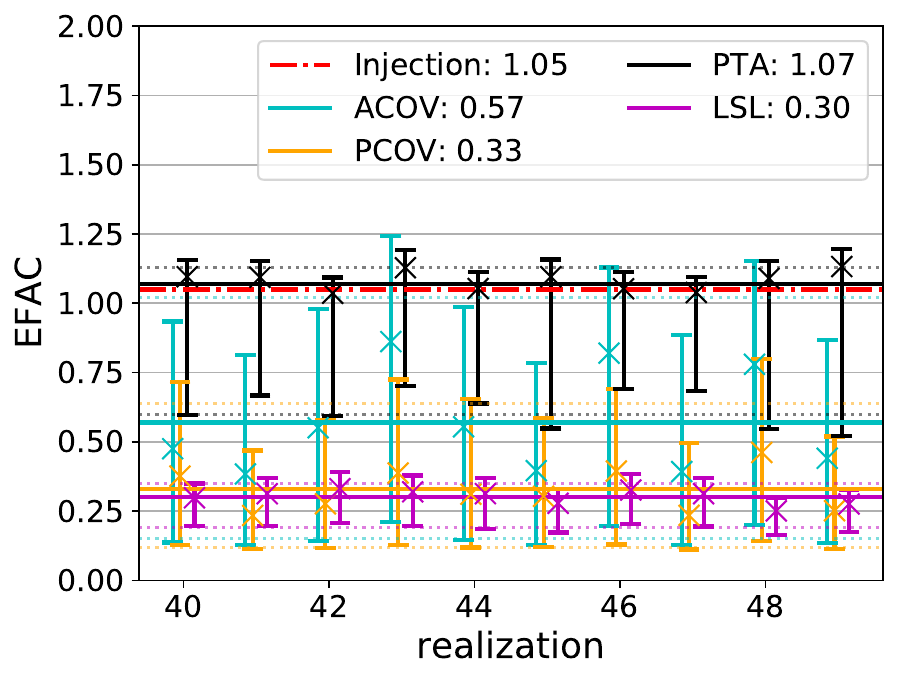}

\caption{Examples of the recovery of the EFAC parameter from one set of realizations of the white noise case (top row), boundary case (middle row) and the red noise case (bottom row): spectral (left column), variance (middle column) and covariance methods (right column). The numbers in the legends represent the average median values of the posterior distributions.}
\label{fig:white_comparison}
\end{figure*}

A different visual aid is given in figure \ref{fig:white_comparison}, which shows the constraints on EFAC from the individual analyses of a subset of 10 sets of realizations.

\subsection{Boundary}

A summary of the constraints on the PTA parameters from the standard PTA/LSL analysis and the clock comparison methods is given in table \ref{table:14_ev}. The posterior distributions for all analysis methods with the recovered spectra and variances can be found in figure \ref{fig:14_recovery}. Examples from analyses of a single simulation set are shown in figure \ref{fig:14_spread}. Despite no injection of a EQUAD value, we still perform the all analyses with all 4 PTA noise parameters. However, the recovered EQUAD posterior distributions are all upper limits as in the white noise case above. Thus, we will omit the EQUAD parameter from our discussion from here on, as the focus will be on the red noise parameters.

\begin{table}
\begin{center}
\def\arraystretch{1.5}
\begin{tabularx}{0.375\textwidth}{c|ccc}
\hline
parameter & ${\rm EFAC}$ & $\gamma_{PTA}$ & $\log_{10} A_{PTA}$ \\
\hline
injection    & $1.05$ & $3$ & $-14$ \\
PTA          & $1.06_{-0.49}^{+0.06}$ & $3.55_{-1.76}^{+2.32}$ & $-14.48_{-1.24}^{+0.75}$ \\
LSL          & $0.23_{-0.07}^{+0.03}$ & $3.24_{-1.10}^{+1.62}$ & $-14.16_{-0.55}^{+0.35}$ \\
PSD          & $1.02_{-0.50}^{+0.09}$ & $3.69_{-2.00}^{+2.59}$ & $-14.75_{-1.75}^{+1.09}$ \\
CS           & $0.42_{-0.18}^{+0.07}$ & $3.55_{-1.19}^{+1.44}$ & $-14.35_{-0.71}^{+0.50}$ \\
AVAR         & $1.04_{-0.52}^{+0.12}$ & $2.78_{-1.07}^{+1.33}$ & $-15.36_{-2.91}^{+1.50}$ \\
PVAR         & $0.92_{-0.44}^{+0.10}$ & $2.76_{-1.13}^{+1.48}$ & $-15.35_{-4.03}^{+1.52}$ \\
ACOV         & $0.30_{-0.16}^{+0.19}$ & $2.99_{-1.21}^{+1.26}$ & $-15.77_{-3.27}^{+1.67}$ \\
PCOV         & $0.25_{-0.13}^{+0.17}$ & $2.97_{-1.31}^{+1.35}$ & $-16.49_{-3.16}^{+2.33}$ \\
\hline
\end{tabularx}
\caption{List of the average median posterior values and average 90\% central region bounds for the 3 constrainable PTA noise parameters from the analyses of 100 realizations in the boundary case with different analysis methods.}
\label{table:14_ev}
\end{center}
\end{table}

\subsubsection{Spectral analysis}

The comparison of the marginalized 1D and 2D posterior distributions of the PTA noise parameter for the boundary case from the spectral analysis can be found in the top left corner plot of figure \ref{fig:14_recovery}. The middle and bottom left panels in figure \ref{fig:14_recovery} show the injected red and white noise along with the recovered PSD and CS respectively.

The red noise signal is not well detectable for the PSD as only the lowest frequency bin in the left middle panel of figure \ref{fig:14_recovery} shows signs of deviating away from white noise. There is a small hint of the red noise. Which is sufficient to put loose constraints on the spectral index $\gamma_{PTA}$ and amplitude $A_{PTA}$ from the PSD method in the top left panel. The constraints on the two PTA red noise parameters are comparable to the standard PTA analysis.

On the other hand, the decrease of the white noise level through the CS improves the detection of the injected red noise and the constraints. The lowest 3 frequency bins contribute to the recovery of the red noise power law, see bottom left panel of figure \ref{fig:14_recovery}. Both the posterior distributions in the top panel of the figure and table \ref{table:14_ev} show that the CS performs a little better with tighter constraints than both the PSD and the standard PTA methods. In the LSL, the signal is clearly detectable and the tightest constraints can be placed in the top left panel.

\subsubsection{Variance analysis}

\begin{figure*}

\includegraphics[width=0.3\linewidth]{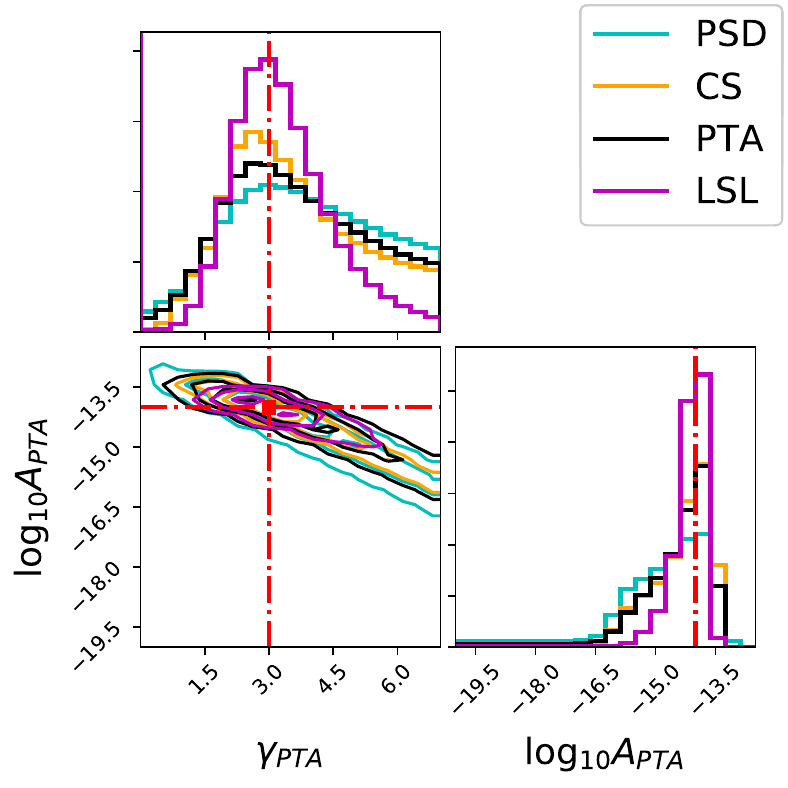} \hspace{0.5cm}
\includegraphics[width=0.3\linewidth]{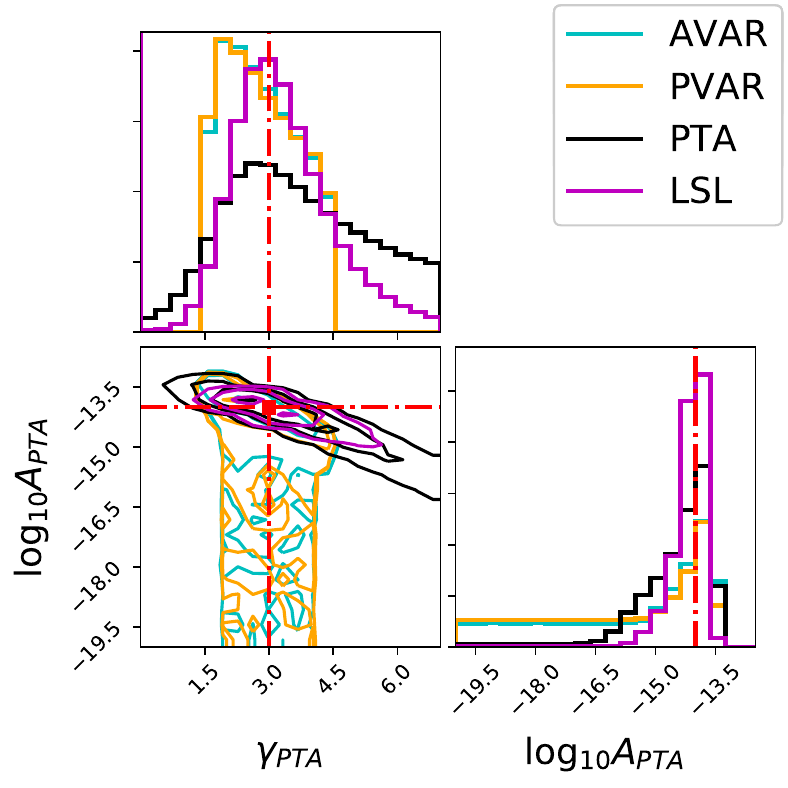} \hspace{0.5cm}
\includegraphics[width=0.3\linewidth]{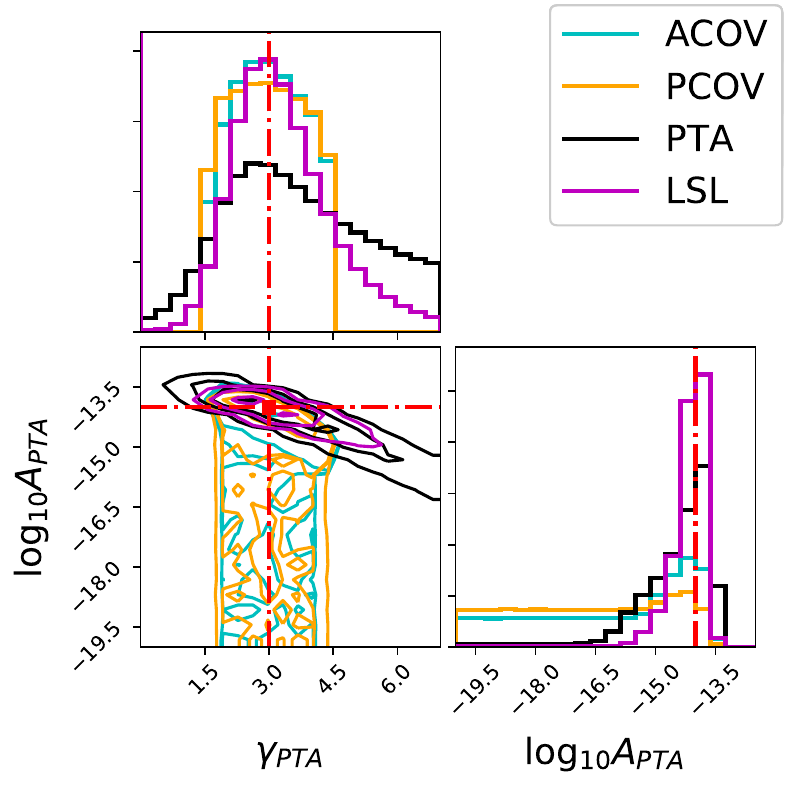}
\\
\includegraphics[width=0.3\linewidth]{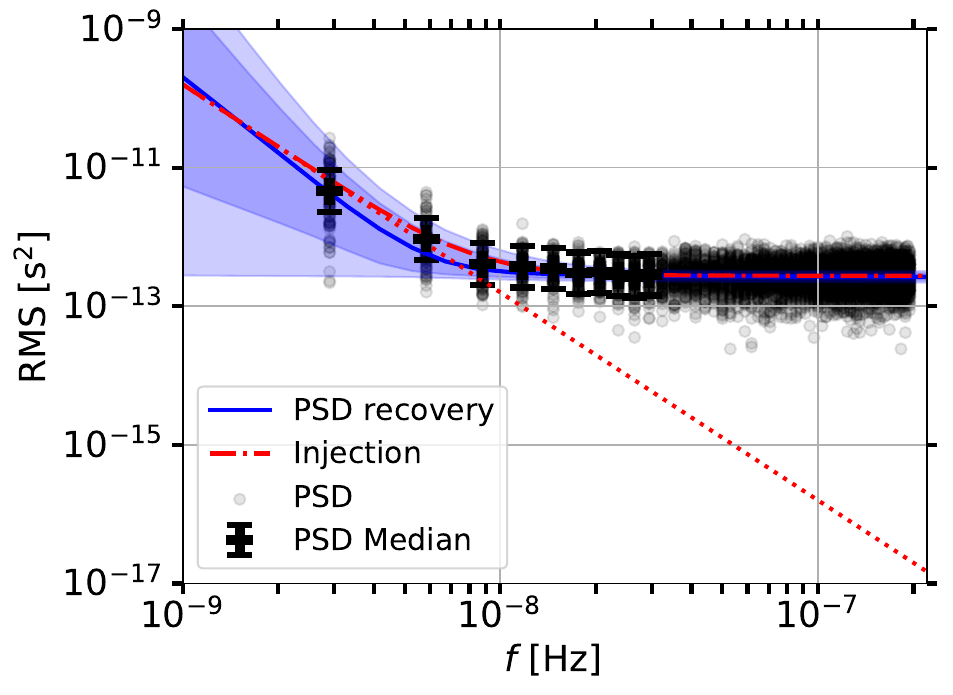} \hspace{0.5cm}
\includegraphics[width=0.3\linewidth]{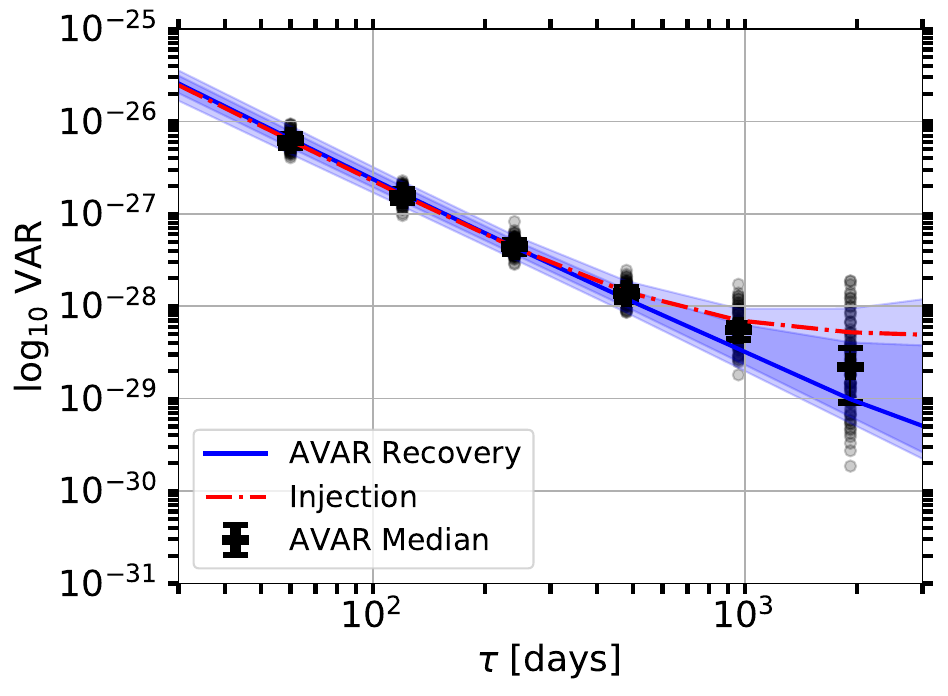} \hspace{0.5cm}
\includegraphics[width=0.3\linewidth]{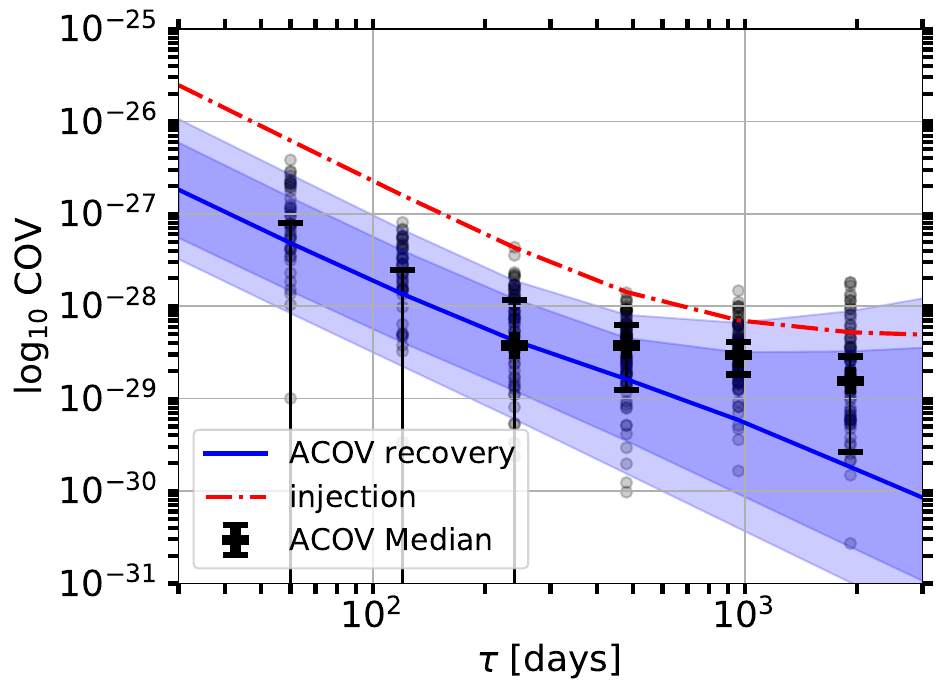}
\\
\includegraphics[width=0.3\linewidth]{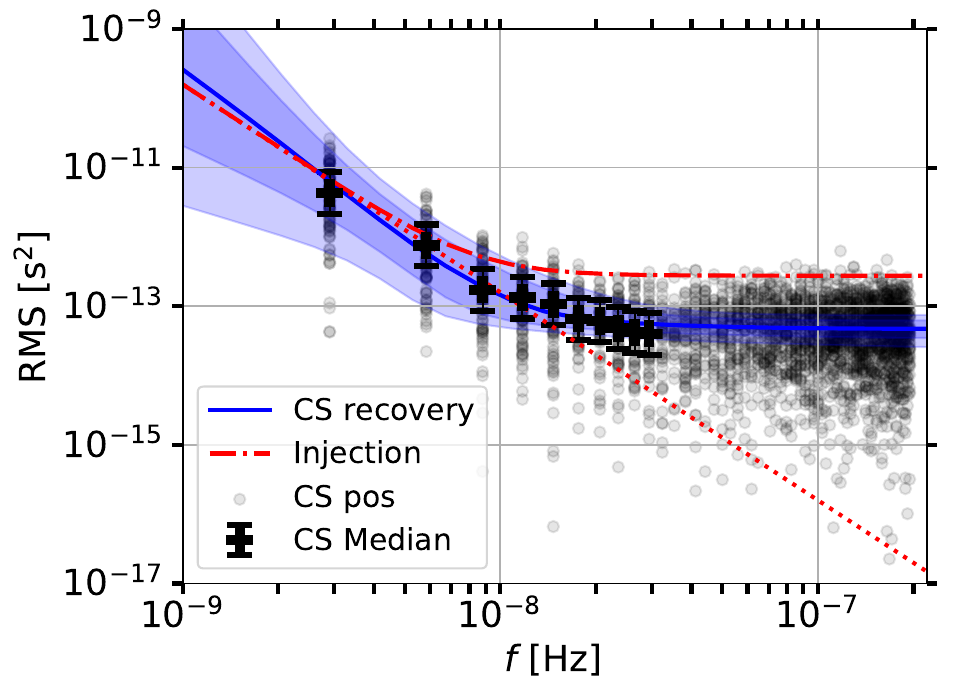} \hspace{0.5cm}
\includegraphics[width=0.3\linewidth]{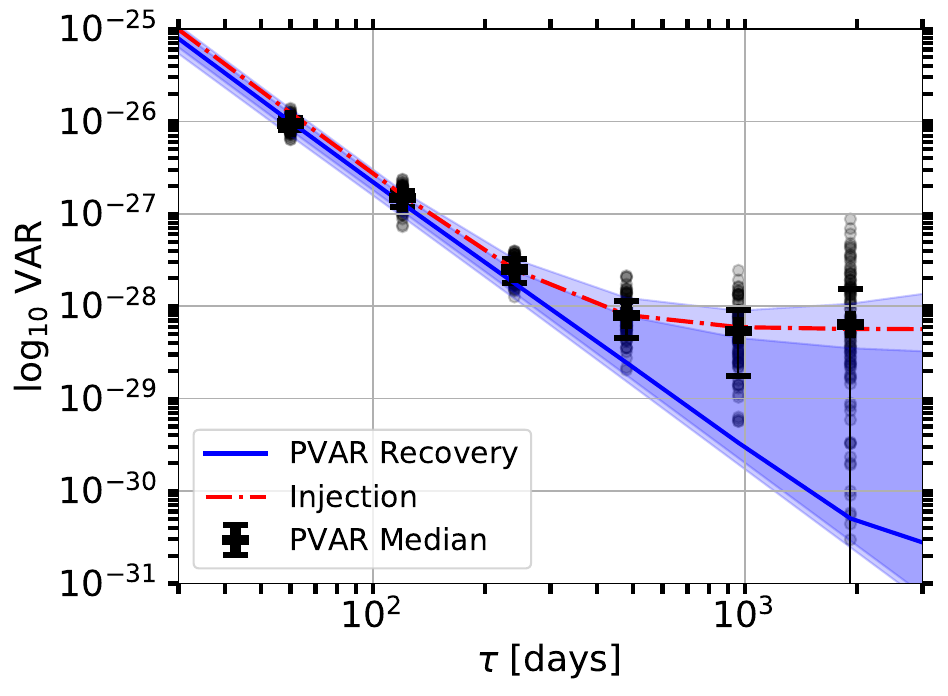} \hspace{0.5cm}
\includegraphics[width=0.3\linewidth]{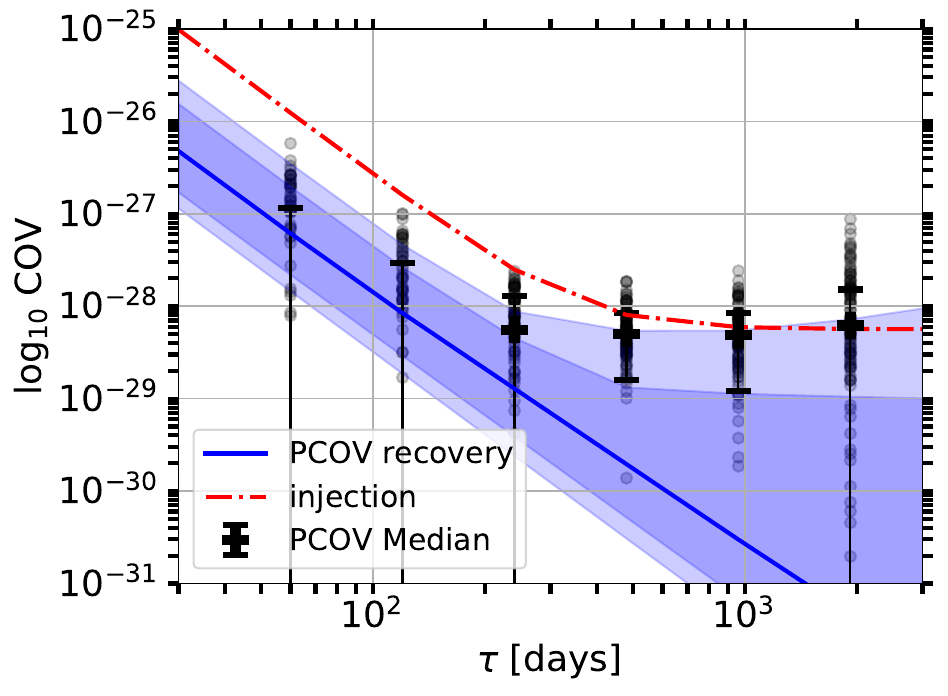}

\caption{The left, middle and right column show results from the spectral, variance and covariance analysis respectively for the boundary case. The top row shows the posterior distribution corner plots. The middle row shows the recovered PSD (left), AVAR (middle) and ACOV (right). The bottom row shows the recovered CS (left), PVAR (middle) and PCOV (right). Each recovery figure contains the values from the 100 realizations in black dots, the injection in red dotted lines and the median values and central 68 and 90\% credible regions in blue bands.}
\label{fig:14_recovery}

\includegraphics[width=0.3\linewidth]{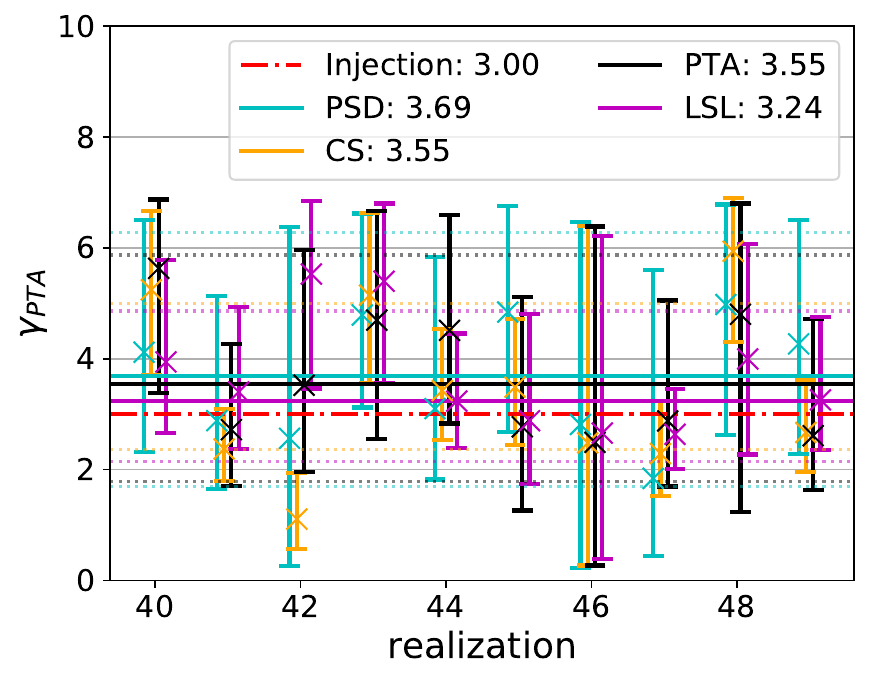} \hspace{0.5cm}
\includegraphics[width=0.3\linewidth]{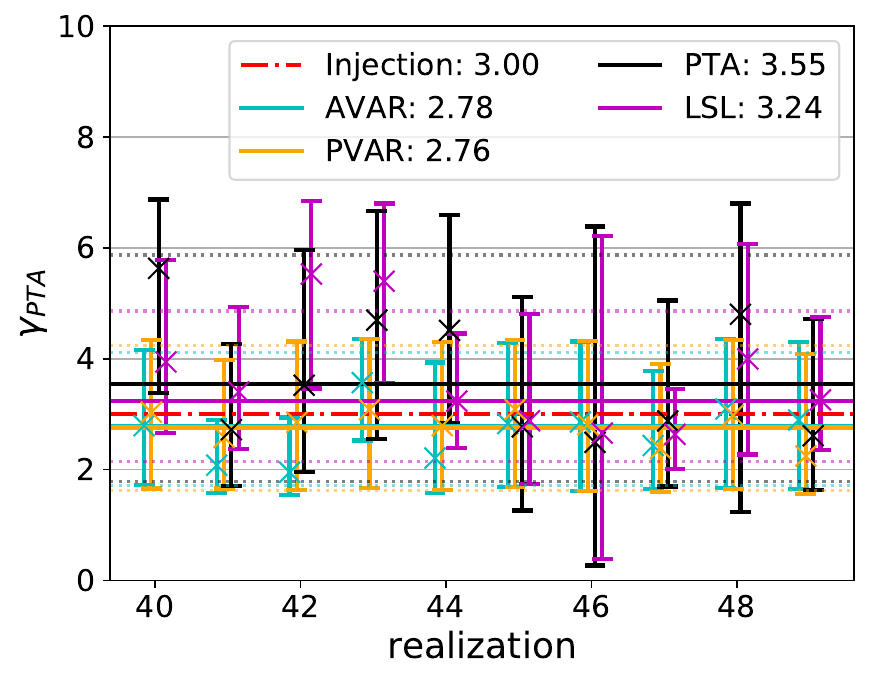} \hspace{0.5cm}
\includegraphics[width=0.3\linewidth]{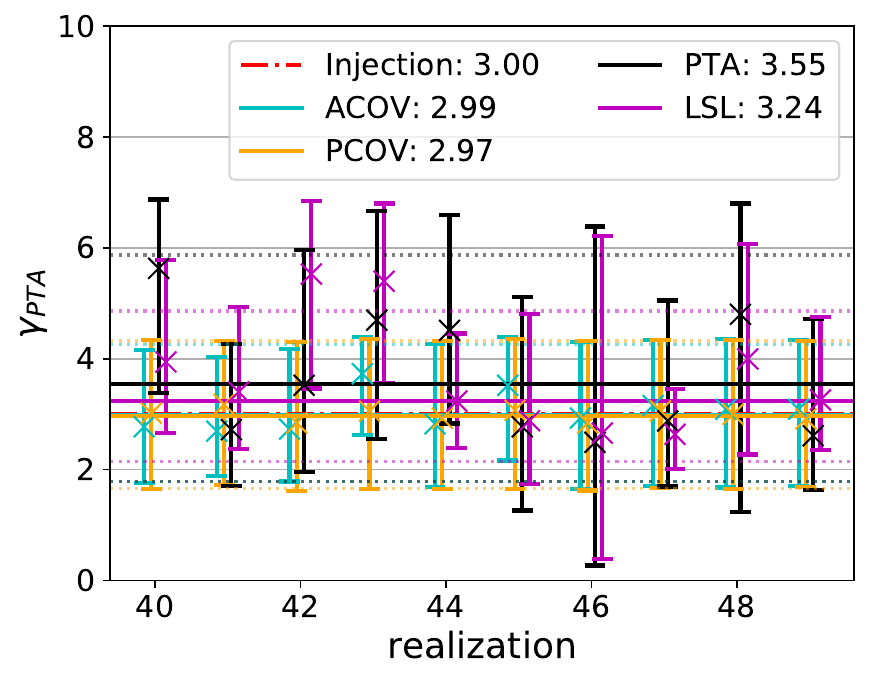}
\\
\includegraphics[width=0.3\linewidth]{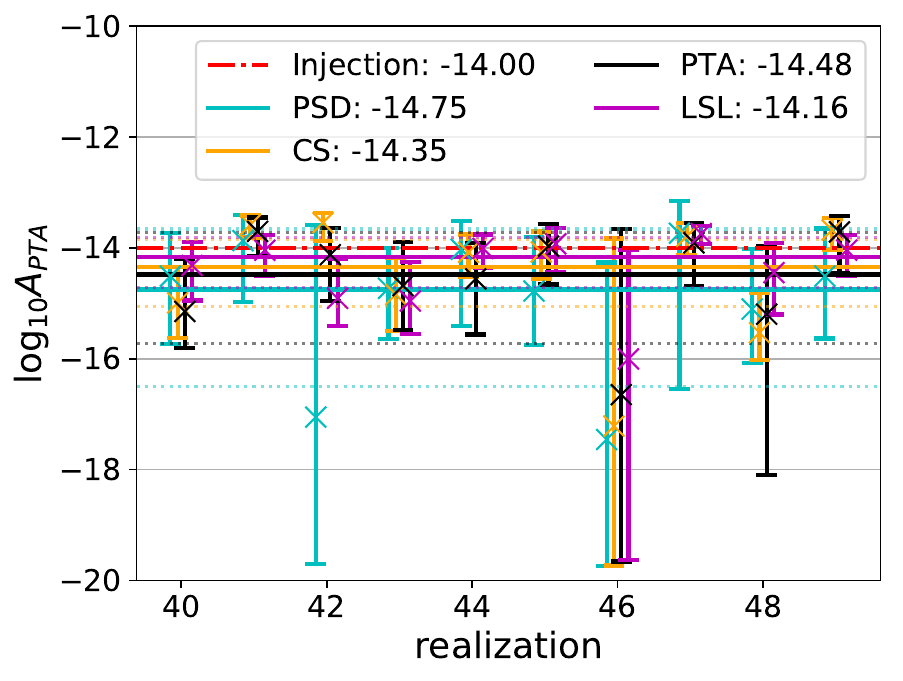} \hspace{0.5cm}
\includegraphics[width=0.3\linewidth]{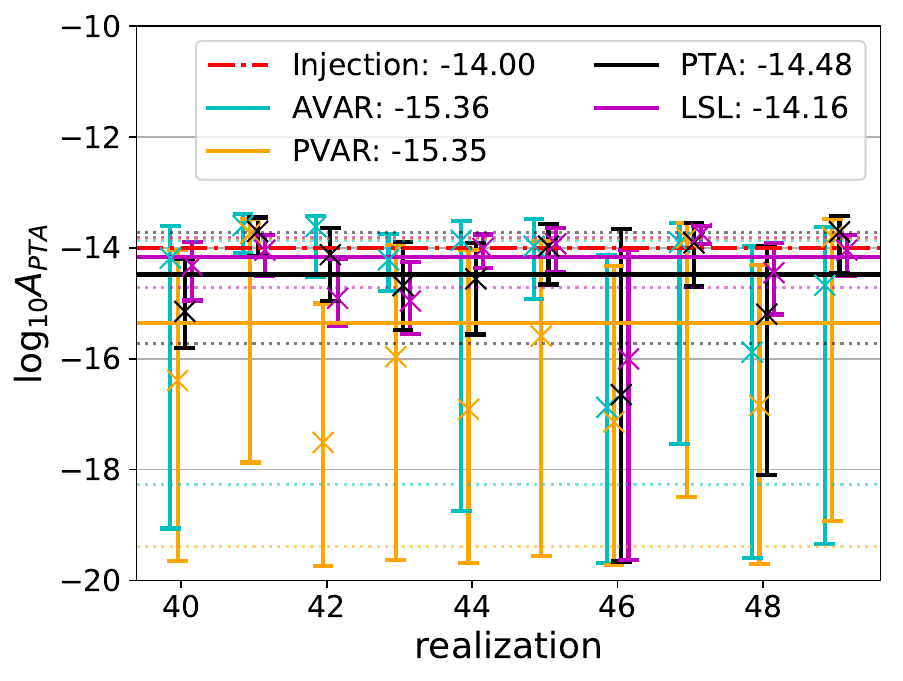} \hspace{0.5cm}
\includegraphics[width=0.3\linewidth]{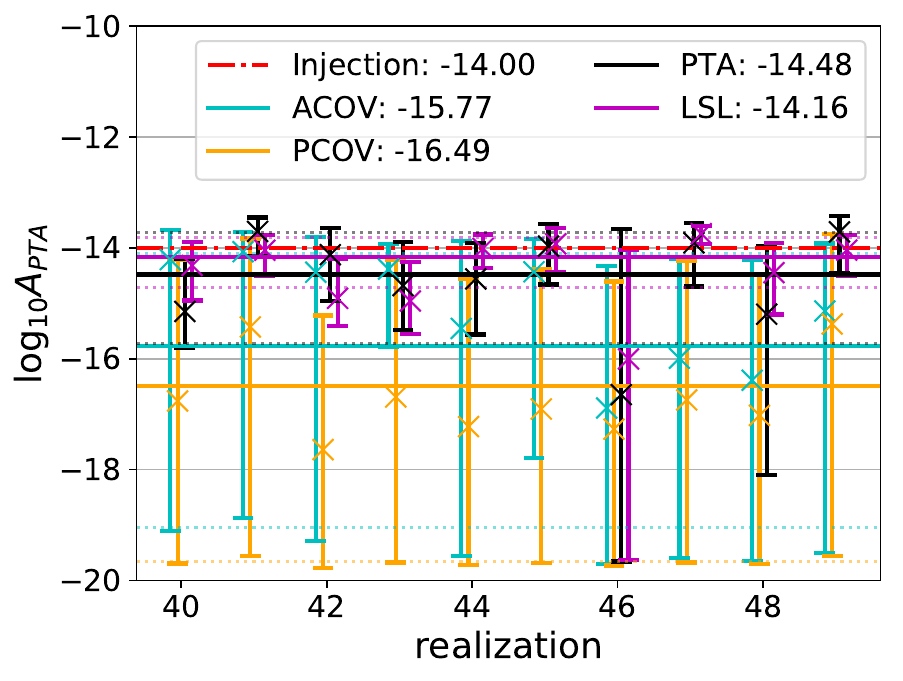}

\caption{Examples of the recovery of the $\gamma_{PTA}$ (top row) and $A_{PTA}$ (bottom row) parameters to central 90\% from one set of realizations of the boundary case: spectral (left column), variance (middle column) and covariance methods (right column). The numbers in the legends represent the average median values of the posterior distributions.}
\label{fig:14_spread}
\end{figure*}

The middle column of figure \ref{fig:14_recovery} shows the comparison red noise parameter corner plot in the top panel and recovered Allan variance (middle panel) and parabolic variance (bottom panel) with the injected red noise. As equation \eqref{eqn:VAR_norm} is only valid for $1.5 \leq \gamma_{PTA} \leq 4.5$, the boundaries are reduced in the top row middle panel of figure \ref{fig:14_recovery}. With this caveat, the red noise spectral index $\gamma_{PTA}$ and amplitude $A_{PTA}$ can still be compared, however the cutoffs at 1.5 and 4.5 limit the comparison and can thus influence the red noise recovery.

The recovered variances in the middle and bottom panels of the middle column of figure \ref{fig:14_recovery} illustrate the fact that the highest octave of $\tau$ with very large uncertainty is unreliable and contributes very little to the overall analysis. Since we are looking at a boundary case, the injections are close or within the uncertainties of the variance computation. Therefore, they do not necessarily coincide with the median of the recovered variance, see also table \ref{table:14_ev}. Additionally, the prior choice allows for a large distribution of small amplitude red noise, producing a median variance close to white noise.

This results in a conservative estimation that there is little red noise in the boundary case with AVAR and PVAR. However, as there a small turn-up in the higher time steps, both provide some evidence of the injected red noise. This can be seen in the top row middle panel corner plot, especially in the $A_{PTA}$ parameter and a small improvement in constraining the spectral index. The PTA analysis is also able to pick up the red noise and performs better than the variances. With the lowest level of white noise, it is no surprise that the LSL analysis performs the best.

\begin{figure*}

\includegraphics[width=0.3\linewidth]{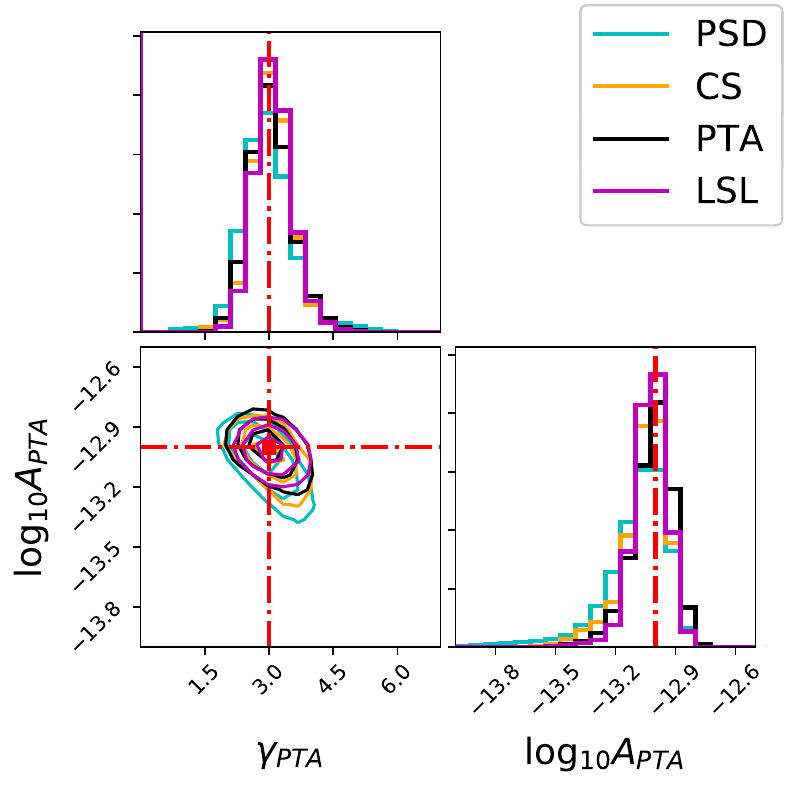} \hspace{0.5cm}
\includegraphics[width=0.3\linewidth]{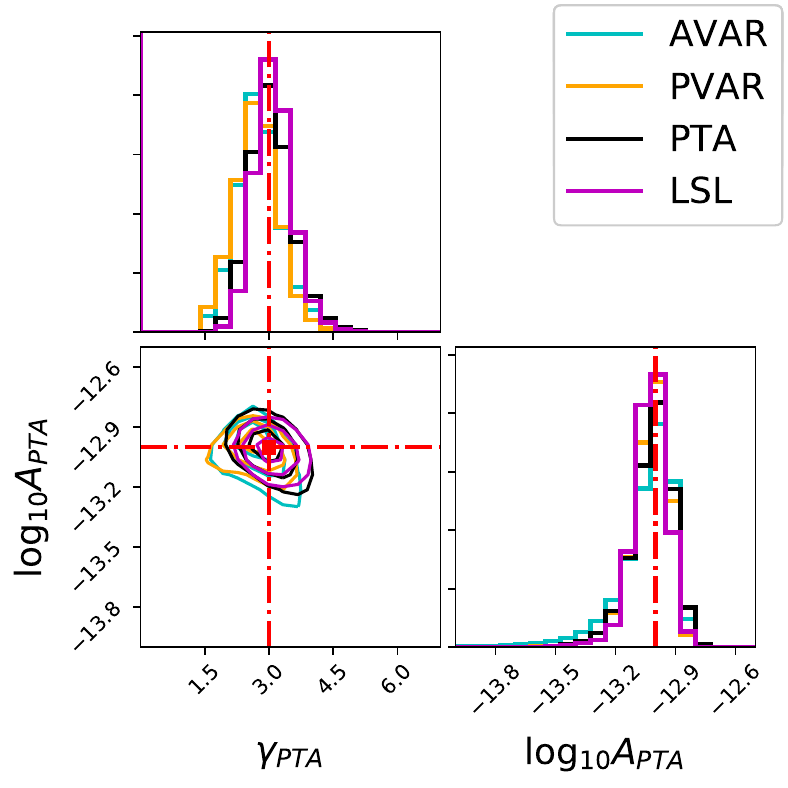} \hspace{0.5cm}
\includegraphics[width=0.3\linewidth]{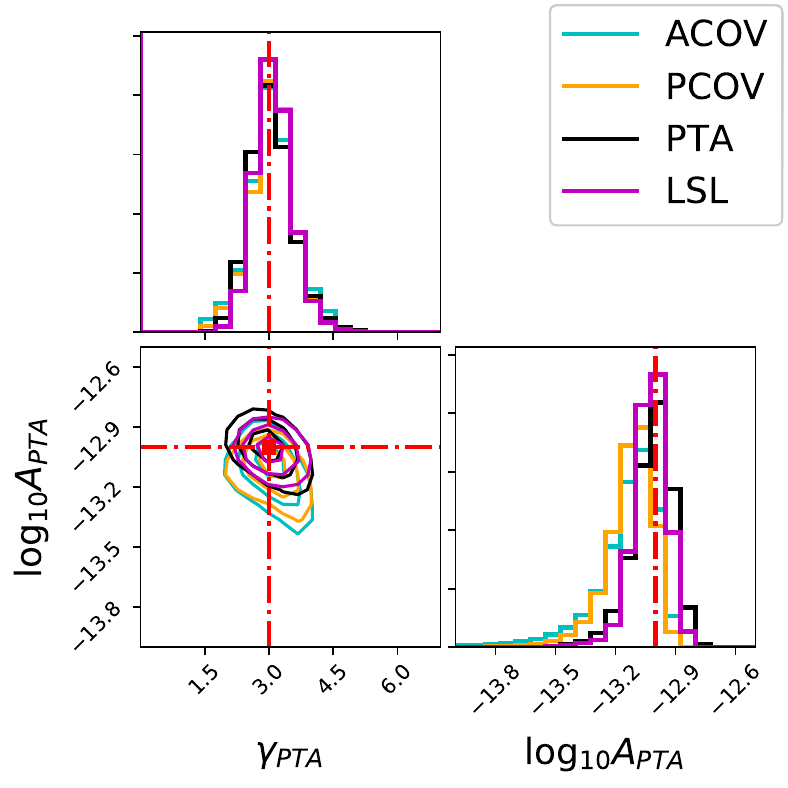}
\\
\includegraphics[width=0.3\linewidth]{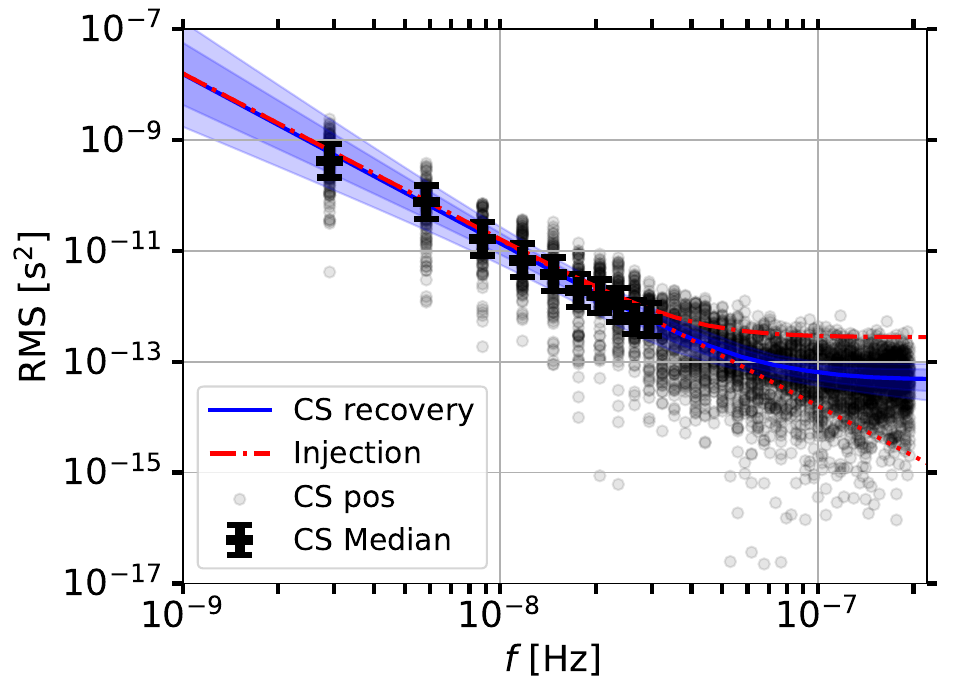} \hspace{0.5cm}
\includegraphics[width=0.3\linewidth]{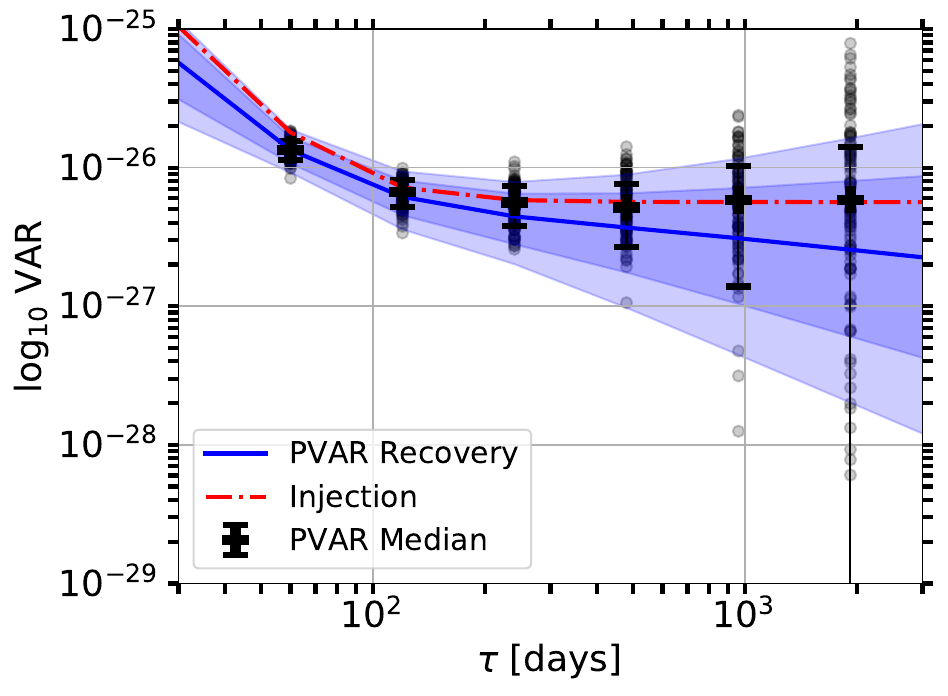} \hspace{0.5cm}
\includegraphics[width=0.3\linewidth]{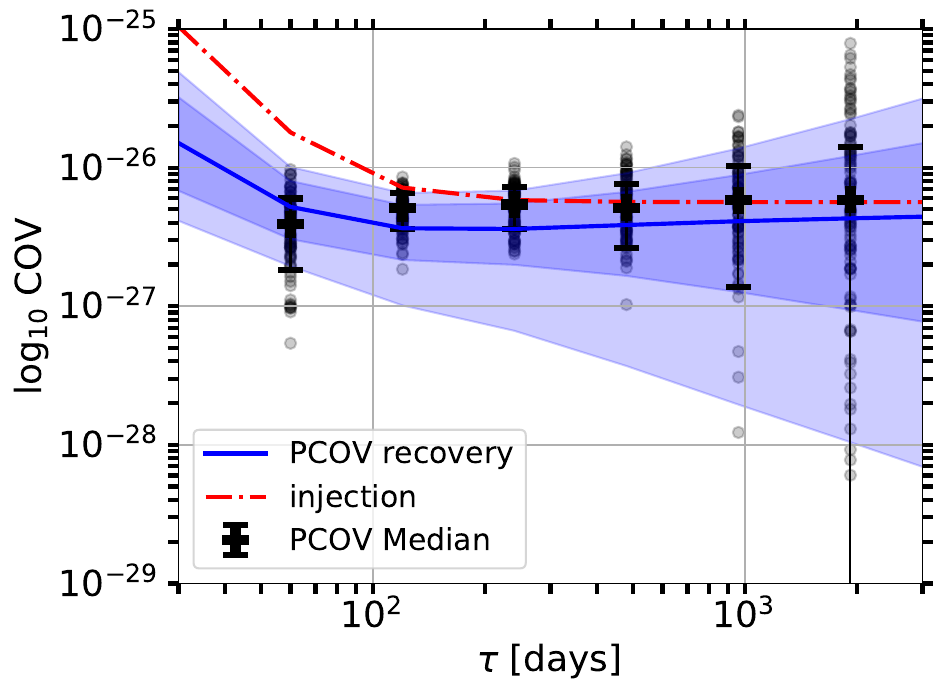}

\caption{Top and bottom rows correspond those in figure \ref{fig:14_recovery} respectively in the same style. However, the results are for the red noise case.}
\label{fig:13_recovery}

\includegraphics[width=0.3\linewidth]{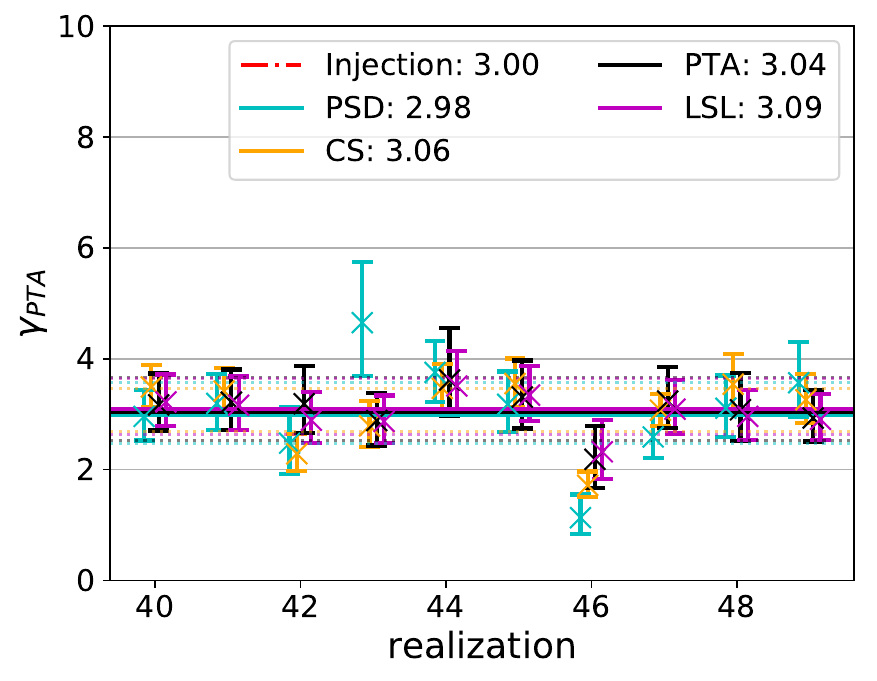} \hspace{0.5cm}
\includegraphics[width=0.3\linewidth]{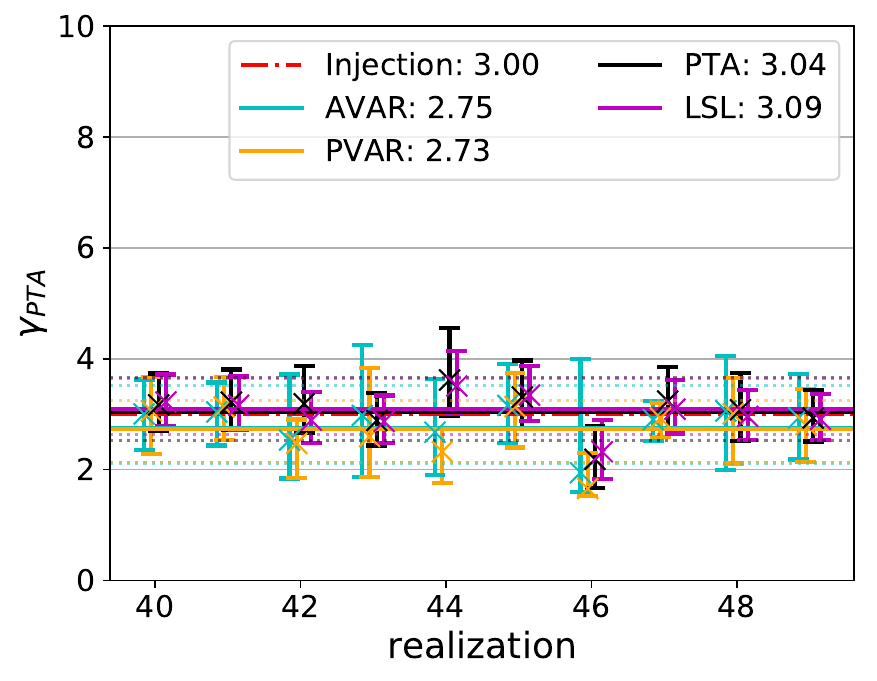} \hspace{0.5cm}
\includegraphics[width=0.3\linewidth]{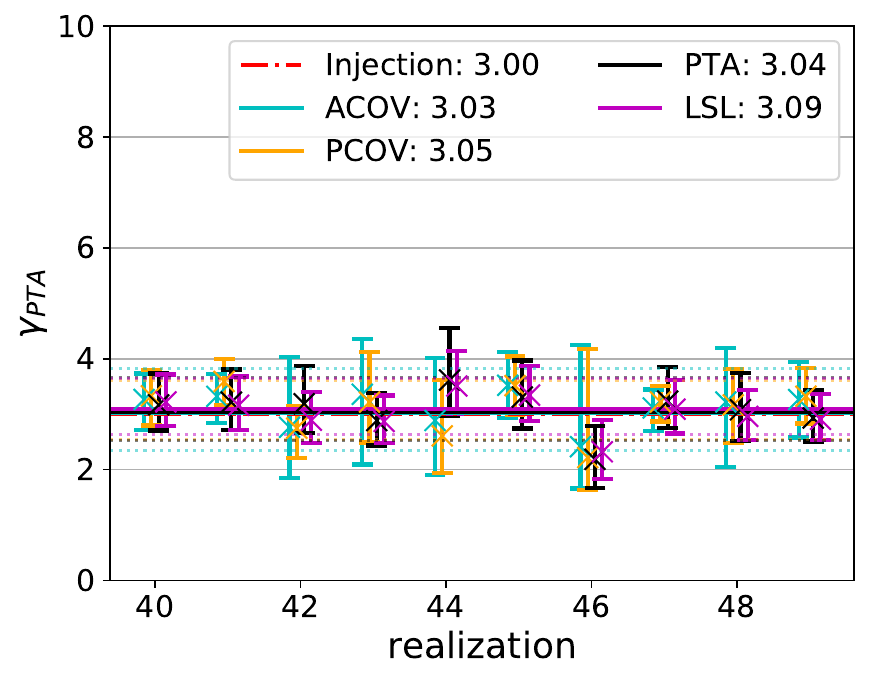}
\\
\includegraphics[width=0.3\linewidth]{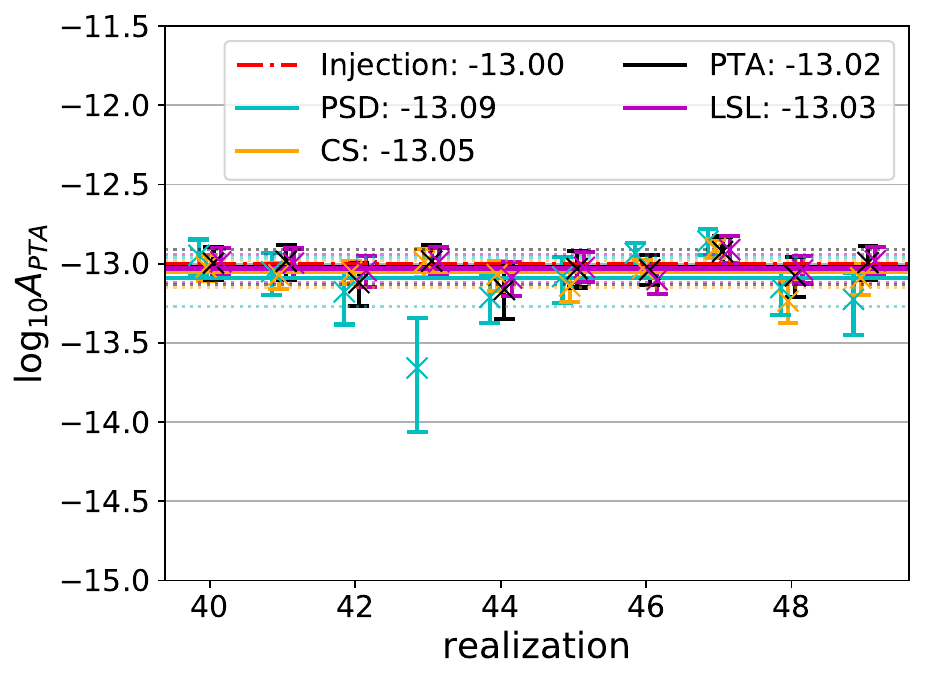} \hspace{0.5cm}
\includegraphics[width=0.3\linewidth]{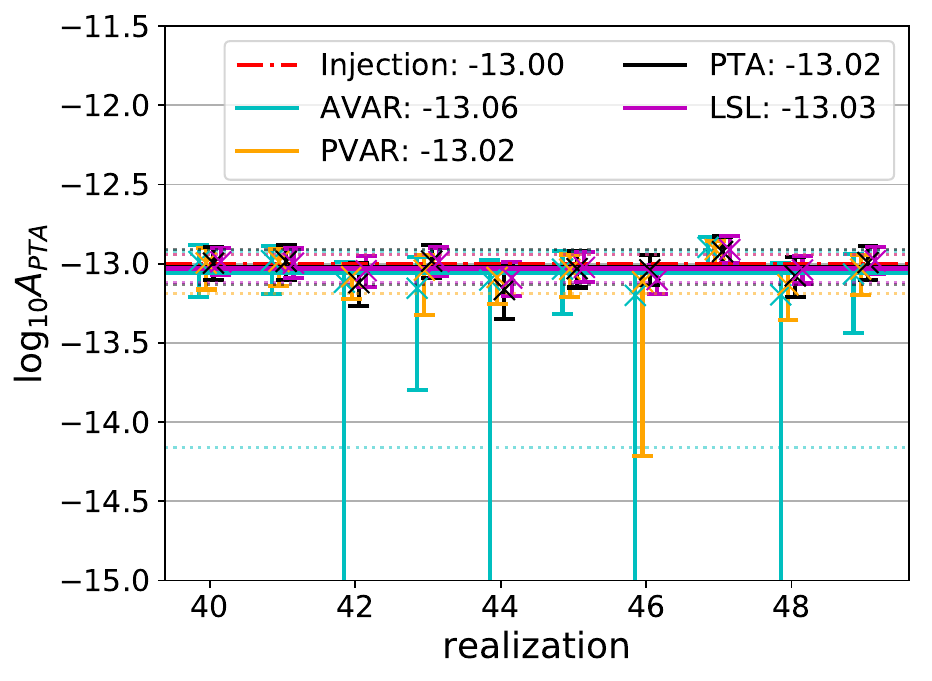} \hspace{0.5cm}
\includegraphics[width=0.3\linewidth]{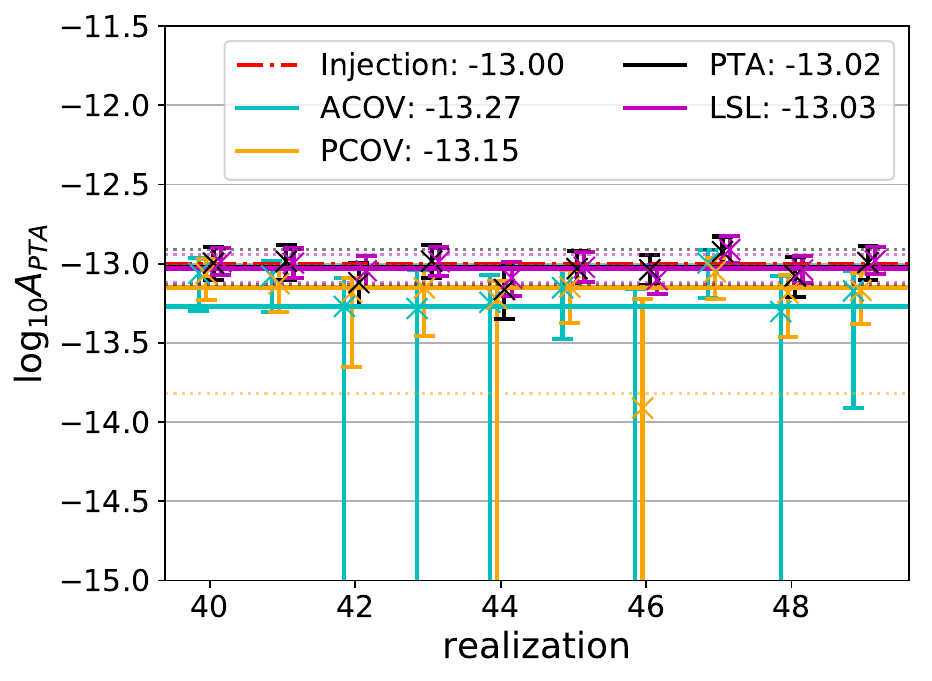}

\caption{Examples of the recovery of the $\gamma_{PTA}$ (top row) and $A_{PTA}$ (bottom row) parameters from one set of realizations of the red noise case: spectral (left column), variance (middle column) and covariance methods (right column). The numbers in the legends represent the median values of the posterior distributions.}
\label{fig:13_spread}
\end{figure*}

\subsubsection{Covariance analysis}

We can further focus on the red noise detection by sacrificing the constraints on the white noise. This is done in a similar way as with the CS by computing the covariances, such that uncorrelated white noise is rejected to allow for the correlated red noise to be more prominent. The right-hand column of figure \ref{fig:14_recovery} shows the posterior distributions, recovered ACOV and PCOV from top to bottom. The same limitation on the spectral index $\gamma_{PTA}$ for variances applies also to the covariances.

Despite the lower white noise from the covariances, the constraints on the red noise parameters $\gamma_{PTA}$ and $A_{PTA}$ are actually worse than their variance counterparts, as it can be seen in the top right panel of figure \ref{fig:14_recovery} and table \ref{table:14_ev}. As there is no correlated white noise, the lowest time steps become upper limits with large uncertainties and do not provide any information on the red noise. This leaves only one or two time steps with information, which is shown in the right column middle and bottom panels. The overall recovery using all time steps thus favours low amplitude white and red noise. Using only time steps with small uncertainties could produce tighter, but less reliable, red noise constraints.

\subsection{Red noise}

\begin{table}
\begin{center}
\def\arraystretch{1.5}
\begin{tabularx}{0.375\textwidth}{c|ccc}
\hline
parameter & ${\rm EFAC}$ & $\gamma_{PTA}$ & $\log_{10} A_{PTA}$ \\
\hline
injection    & $1.05$ & $3$ & $-13$ \\
PTA          & $1.07_{-0.47}^{+0.06}$ & $3.04_{-0.52}^{+0.62}$ & $-13.02_{-0.11}^{+0.11}$ \\
LSL          & $0.30_{-0.11}^{+0.05}$ & $3.09_{-0.46}^{+0.55}$ & $-13.03_{-0.09}^{+0.09}$ \\
PSD          & $1.01_{-0.50}^{+0.12}$ & $2.98_{-0.51}^{+0.59}$ & $-13.09_{-0.18}^{+0.13}$ \\
CS           & $0.42_{-0.18}^{+0.08}$ & $3.06_{-0.38}^{+0.41}$ & $-13.05_{-0.10}^{+0.07}$ \\
AVAR         & $0.90_{-0.69}^{+0.46}$ & $2.75_{-0.63}^{+0.77}$ & $-13.06_{-1.10}^{+0.14}$ \\
PVAR         & $0.68_{-0.48}^{+0.34}$ & $2.73_{-0.60}^{+0.52}$ & $-13.02_{-0.17}^{+0.08}$ \\
ACOV         & $0.57_{-0.42}^{+0.45}$ & $3.03_{-0.69}^{+0.79}$ & $-13.27_{-2.24}^{+0.26}$ \\
PCOV         & $0.33_{-0.21}^{+0.31}$ & $3.05_{-0.50}^{+0.55}$ & $-13.15_{-0.67}^{+0.12}$ \\
\hline
\end{tabularx}
\caption{List of the average median posterior values and average 90\% central region bounds for the 3 constrainable PTA noise parameters from the analyses of 100 realizations in the red noise case with different analysis methods.}
\label{table:13_ev}
\end{center}
\end{table}

The strong red noise case is of particular interest as it shows how well the different methods can in principle divide different red noise realizations in the large-signal regime. An overview on the constraints on the PTA parameters for the red noise case is given by table \ref{table:13_ev} for the standard PTA/LSL and clock metrology analysis. The overall posterior distributions for all methods are shown in figure \ref{fig:13_recovery}. The results of some example analyses of a single simulation set can be found in figure \ref{fig:13_spread}.

In general, all methods can detect the injected red noise very well. As the red noise is very large, there is very little difference in the constraints on the red noise parameters $\gamma_{PTA}$ and $\log_{10} A_{PTA}$ between the PTA and the LSL analysis. Both are centered around the injected values with tight constraints, see top panels of figure \ref{fig:13_recovery}. So do the spectral method recoveries, although with a very small tail, which is longer for the PSD compared to the CS. The uncertainties in the spectral methods allow for small values even at the lowest frequencies. This leads to a tail of lower amplitudes being recovered, see table \ref{table:13_ev}. Both variances and covariances perform very similarly to the PTA/LSL analysis in terms of the tightness of the constraints on the red noise parameters, especially PVAR and PCOV are comparable to the LSL analysis. This is due to the fact that the higher time steps are dominated by red noise and have therefore very similar VAR/COV values. Additionally, the absolute uncertainties $\sigma_i$ are the same for both variances and covariances.

Consequently, the covariances at the lowest $\tau$ are not as well defined as the variances for the same time steps. These large relative uncertainties and the lower COV values translate into a small bias for the red noise to have steeper spectral indices $\gamma_{PTA}$ than with the variance methods. On the other hand, the variance methods are still a little bit impacted by the choice of priors, see table \ref{table:13_ev}. This can be seen in the bottom middle panel of figure \ref{fig:13_recovery}, where the median recovered PVAR is lower than the median PVAR values of the 100 realizations. The same is also true for AVAR recovery.

Table \ref{table:13_ev} also shows that both AVAR and ACOV have a longer tail towards lower red noise amplitudes $\log_{10} A_{PTA}$ when compared to PVAR and PCOV. This can be the advantage of the parabolic (co)variance transitioning from a spectral index of $\beta_{VAR} = -3$ for white noise to $\gamma_{VAR} = 0$ for the injected red noise. In comparison the Allan (co)variance has a transition of one order less, ie. from $\beta_{VAR} = -2$ to $\gamma_{VAR} = 0$.


\section{Conclusions}
\label{sec:conclusions}

We have shown that clock comparison methods, such as the cross spectrum and variances, can be applied to pulsar timing observations. Initial tests show that these methods produce slightly worse, sometimes comparable constraints on the pulsar noise parameters when compared to the standard PTA and optimal LSL analysis. Both perform very well in recovering the injected white noise, regardless of the level of red noise. The red noise recovery itself is broadly consistent with the injections, with different realizations giving similar constraints. In general, the metrology clock comparison methods perform well in the pure white noise case, slightly worse in the boundary case and comparable in the strong red noise case. The spread of recoveries between different realizations is also larger than the comparison PTA/LSL analysis.

The PSD method produces slightly broader parameter posterior distributions compared to those from the PTA analysis. The CS allows for a long tail of low values, including negative values. This means a reduction of white noise to almost zero at high frequencies, but including these negative values will skew the red noise recovery. Using Gaussian distributions also decreases the possibility for low CS values and thus give slightly more optimistic constraints for $A_{PTA}$ and $\gamma_{PTA}$. Thus, the CS method performs slightly better than the PTA analysis, but still a bit worse than the LSL analysis. Since the LSL's white noise level is only a fifth of that of the PTA observations, it can put the tighest constraints on the red noise.

Similar constraints can be put on the red noise parameters for both the variances and covariances. The (co)variance methods perform particularly badly in recovering the white noise in the red noise case. By design, ACOV and PCOV should reject uncorrelated white noise and only leave the correlated red noise signal. Therefore, no useful information can be gained for the white noise, while some extra power may be pushed into the red noise. This lack of information may limit the performance of the covariances in precisely constraining the red noise and manifests as a small bias in $\gamma_{PTA}$ or a long tail of low red noise amplitudes $A_{PTA}$.

The clock comparison methods complement the standard PTA analysis by giving independent constraints on the noise parameters. This can be especially valuable for certain realizations, when the different analysis methods do not converge to the same values.

We continue the study and apply the methods described to the observations from the LEAP project, where we deal with the challenges from real pulsar timing data. These include unevenly sampled data points, different levels of white noise between RTs and observations and uncertainties in the computation of the spectral densities and (co)variances.


\section*{Data availability}
The simulated residual series and posteriors can be obtained from the authors upon request.

\section*{Acknowledgements}
This work is funded by the ANR Programme d'Investissement d'Avenir (PIA) under the FIRST-TF network (ANR-10-LABX-48-01) project and the Oscillator IMP project (ANR-11-EQPX-0033-OSC-IMP) and the EUR EIPHI Graduate School (ANR-17-EURE-00002), and by grants from the R\'{e}gion Bourgogne Franche Comt\'{e} intended to support the PIA.

The Nan{\c c}ay radio Observatory is operated by the Paris Observatory, associated to the French Centre National de la Recherche Scientifique (CNRS). We acknowledge financial support from "Programme National de Cosmologie and Galaxies" (PNCG) and "Programme National Hautes Energies" (PNHE) funded by CNRS/INSU-IN2P3-INP, CEA and CNES, France.

We acknowledge the support of our colleagues in the Large European Array for Pulsars and European Pulsar Timing Array collaboration.

\bibliographystyle{mnras}
\bibliography{Ref-short,References,Ref-Rubiola,Ref-Chen}

\appendix

\section{Estimates and weights of the cross spectrum}
\label{sec:weights_sd}

For a given frequency $f_0$, we consider the noise level $S_w$ from uncorrelated observational noise of the RTs and the signal level $S_r$ from the pulsar red noise, see equation \eqref{eqn:PTA_SD}. We perform a large Monte-Carlo simulation with $10^7$ realizations at the given frequency with varying levels of noise and signal.

\begin{figure}
\centering
\includegraphics[width=\linewidth]{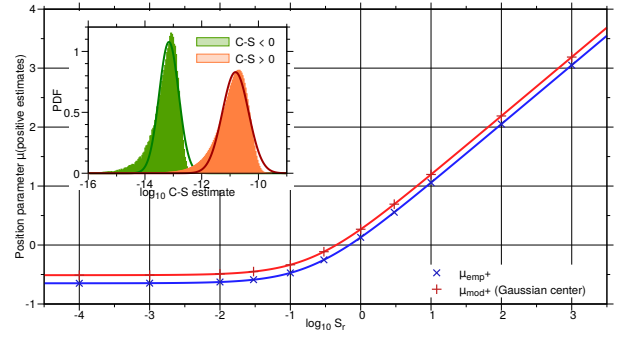}
\caption{Dependence on $S_r$ of the log mean (in blue) and of the Gaussian center (in red) for positive CS estimates when $S_w=1$. Inset: Gaussian fit of the histograms of the log of the CS estimates ($S_w=3\cdot 10^{-13}$, $S_r=10^{-11}$). \label{fig:log-mean}}
\end{figure}

The large simulation allows for the evaluation of the estimates of the CS. We determine empirically the statistics and conclude that the most reliable representation can be obtained from the logarithmic CS estimates distinguishing between positive and negative estimates. The empirical distributions are described by two parameters: $\mu_{emp}$ and $\sigma_{emp}$ are the mean and variance of the log of the CS estimates respectively. Despite the asymmetry of the probability density functions we choose to model them with Gaussians with mean $\mu_{mod}$ and variance $\sigma_{mod}$ and find the best fit values to both positive and negative CS histograms (see inset of figure \ref{fig:log-mean}).

\begin{table}
\begin{center}
\def\arraystretch{1.5}
\begin{tabularx}{0.5\textwidth}{c|c|c}
\hline
Parameter & Negative estimates & Positive estimates \\
\hline
\hline
&&\\
$\mu_{emp}$ & $\displaystyle \log_{10}\left(0.1949S_w\frac{S_w+3.876S_r}{S_w+4.552S_r}
\right)$ & $\displaystyle \log_{10}\left[0.2245\left(S_w+5S_r\right)\right]$ \\
&&\\
$\mu_{mod}$ &$\displaystyle \log_{10}\left(0.2665S_w\frac{S_w+3.882S_r}{S_w+4.404S_r}\right)$ & $\displaystyle \log_{10}\left[0.3082\left(S_w+5S_r\right)\right]$\\
&&\\
$\sigma_{emp}$ &$\displaystyle \log_{10}\left(1.576\frac{S_w+4.147S_r}{S_w+3.833S_r}\right)$&$\displaystyle \log_{10}\left(2.045\right)=0.3106$ \\
&&\\
$\sigma_{mod}$ &$\displaystyle \log_{10}\left(1.251\frac{S_w+3.835S_r}{S_w+3.640S_r}\right)$&$\displaystyle \log_{10}\left(1.635\right)=0.2136$\\
\hline
\end{tabularx}
\caption{Dependence of the position and width parameters of the CS estimates on $S_w$ and $S_r$.\label{tab:results}}
\end{center}
\end{table}

By varying the noise and signal levels we obtain numerical expressions for the 4 parameters as a function of $S_w$ and $S_r$. They can be found in table \ref{tab:results}. The factor $(S_w+5S_r)$ shows that we can retrieve the correct number of degrees of freedom, ie. 5, in a purely empirical way.

The important results are the following:
\begin{enumerate}
\item the width parameters $\sigma_{emp,mod}$ can be considered as constants
\item the position parameters for negative estimates $\mu_{emp,mod}^-$ only depend on $S_w$
\item the position parameters for positive estimates $\mu_{emp,mod}^+$ depend on $S_w$ as well as $S_r$
\item the shift between $\mu_{emp}$ and $\mu_{mod}$ is constant and can be evaluated from the $10^7$ realizations to be $\log_{10}(1.373)=0.138$. 
\end{enumerate}

This shift between simulation and model can be explained by the long tail towards lower numbers of the empirical distributions which decrease $\mu_{emp}$ compared to $\mu_{mod}$.

The (almost) constancy of the width parameters confirms that the log of the CS estimates is a reliable representation. Moreover, the position parameters of the negative estimates do not depend on $S_r$. This implies that the negative estimates do not carry any information about the signal level and should not been taken into account. The variation of the signal and noise levels makes $f_0$ representative for any frequency. Additionally, since the PSD is similar to the positive CS estimator, we can apply the same Gaussian model.

\section{Weights and unbiasing of the variances}
\label{sec:weights_var}

The uncertainty $\sigma_i$ on each time step $i$, whether variances or covariances, can be calculated from the average values of the variances $VAR$
\begin{equation}
\sigma_i^2 = 2 VAR^2 / \nu ,
\label{eqn:VAR_unc}
\end{equation}
where $\nu$ are the effective degrees of freedom, whose expressions depend on the type of noise. The expressions used in the analysis for AVAR can be found in \cite{Lesage-1973-TIM}: for white phase modulation ($\gamma_{PTA}=0$)
\begin{equation}
\nu \approx \frac{(N+1)(N-2m)}{2(N-m)}
\end{equation}
and flicker frequency modulation ($\gamma_{PTA}=3$):
\begin{equation}
\nu \approx \frac{5N^2}{4m(N+3m)} .
\end{equation}

The equation for PVAR is given in \cite{2020arXiv200513631V} as
\begin{equation}
\nu \approx \frac{35}{Am/M-B(m/M)^2},
\end{equation}
where $A$ and $B$ are constants which depend on the type of the modulation. They are given as $A = 23$ for $\gamma_{PTA}=0$ and $A \approx 28$ for $\gamma_{PTA}=3$, whilst $B\approx 12$ for all spectral indices. As pointed out in \cite{2020arXiv200513631V}, the formula is unreliable for the highest $\tau$. However, from the theory we know that $\nu=1$ at the largest time step for PVAR. We also notice a mixture of white and red noise for our realizations, especially for the boundary case. Consequently, we approximate the degrees of freedom for the boundary case using those from the white noise case as a conservative approach. The degrees of freedom used for our analysis can be found in tables \ref{table:13_weights} and \ref{table:15_weights} and are the same for the covariances.

Finally, \cite{6174199} have shown that the largest time steps are statistically biased towards lower VAR/COV values. The factors that need to be applied for any time step $\tau$ with its degrees of freedom $\nu$ can be found in table II in \cite{6174199}. We also show the factors used in this paper in tables \ref{table:13_weights} and \ref{table:15_weights}.

\begin{table}
\begin{center}
\def\arraystretch{1.5}
\begin{tabularx}{0.45\textwidth}{c|cccccc}
\hline
$\tau$ & 60 & 120 & 240 & 480 & 960 & 1920 \\
\hline \hline
assumed $\gamma_{PTA}$    & $3$ & $3$ & $3$ & $3$ & $3$ & $3$ \\
\hline
AVAR $\nu$    & $80$ & $38$ & $18$ & $7.7$ & $3.0$ & $1.1$ \\
PVAR $\nu$    & $82$ & $40$ & $19$ & $8.6$ & $3.4$ & $1$ \\
\hline
AVAR unbiasing factor  & $1$ & $1$ & $1$ & $1.14$ & $1.45$ & $3.56$ \\
PVAR unbiasing factor  & $1$ & $1$ & $1$ & $1.12$ & $1.45$ & $3.56$ \\
\hline
\end{tabularx}
\caption{Effective degrees of freedom and unbiasing factors for the variances in the red noise case.}
\label{table:13_weights}
\end{center}
\end{table}

\begin{table}
\begin{center}
\def\arraystretch{1.5}
\begin{tabularx}{0.45\textwidth}{c|cccccc}
\hline
$\tau$ & 60 & 120 & 240 & 480 & 960 & 1920 \\
\hline \hline
assumed $\gamma_{PTA}$    & $0$ & $0$ & $0$ & $0$ & $0$ & $0$ \\
\hline
AVAR $\nu$    & $66$ & $65$ & $63$ & $58$ & $46$ & $5.8$ \\
PVAR $\nu$    & $100$ & $49$ & $23$ & $11$ & $4.4$ & $1$ \\
\hline
AVAR unbiasing factor  & $1$ & $1$ & $1$ & $1$ & $1$ & $1.19$ \\
PVAR unbiasing factor  & $1$ & $1$ & $1$ & $1.1$ & $1.31$ & $3.56$ \\
\hline
\end{tabularx}
\caption{Effective degrees of freedom and unbiasing factors for the variances in the white noise and boundary case.}
\label{table:15_weights}
\end{center}
\end{table}

\label{lastpage}

\end{document}